\documentclass[11pt]{article}
\usepackage{comment}
\usepackage{lipsum}
\usepackage{booktabs}
\usepackage{array}
\usepackage{threeparttable}
\usepackage{amsthm}

\newtheorem{proposition}{Proposition}

\usepackage{graphicx}
\usepackage[notablist,nofiglist]{endfloat}

\usepackage[utf8]{inputenc}
\usepackage{url}
\usepackage{setspace}
\usepackage{palatino}
\usepackage{float}
\usepackage{titling}
\usepackage{multirow}
\usepackage{lscape}
\usepackage{amsmath}
\usepackage{amssymb}
\usepackage{subcaption}
\usepackage[a4paper, total={6in, 9.5in}]{geometry}
\fontfamily{ppl}\selectfont
\usepackage[american]{babel}
\addto\captionsamerican{}

\usepackage[authoryear,round]{natbib}
\usepackage{booktabs,makecell,adjustbox}
\usepackage{tikz}
\usetikzlibrary{positioning,fit,calc,arrows.meta,backgrounds}
\usepackage{adjustbox}
\pdfoutput=1
\tikzset{
  font=\small,
  block/.style   = {draw, rounded corners=2pt, line width=0.5pt, fill=white,
                    inner sep=6pt, align=center},
  tinyblock/.style = {block, inner sep=4pt},
  groupbox/.style = {draw, rounded corners=3pt, line width=0.7pt, inner sep=6pt},
  arrow/.style   = {-{Stealth[length=3mm]}, line width=0.6pt},
  darr/.style    = {arrow, dashed},
  edgelabel/.style = {
    pos=.56, sloped, allow upside down,
    inner sep=1.2pt,
    font=\scriptsize,
    text width=24mm,
    align=center
  }
}

\title{\textbf{Hope, Signals, and Silicon: A Game-Theoretic Model of the Pre-Doctoral Academic Labor Market in the Age of AI}\thanks{I am deeply grateful to Editor Ben R. Martin, Associate Editor Daniele Rotolo, and two anonymous reviewers for the exceptional care, time, and intellectual generosity they devoted to this manuscript. I am particularly indebted to the two reviewers, whose thoughtful and incisive reports prompted fundamental, direction-changing revisions to the paper's methodology and to the way its contribution is framed. Their comments led me to adopt a more appropriate and well-justified methodological approach, focus the research question much more sharply, and clarify both the problem the paper addresses and the contribution it seeks to make. Their guidance substantially reshaped the direction of the research and greatly strengthened its underlying ideas, analysis, and exposition. I sincerely appreciate their intellectual generosity and the care with which they helped bring the paper to its present form. All remaining errors are my own.}}
\author{Shaohui Wang\\
J. Mack Robinson College of Business\\
Georgia State University\\
\texttt{swang83@gsu.edu}}
\date{}

\begin{document}
{\setstretch{.8}
\maketitle
\vspace{-1.2em}
\begin{center}
\small
The version of record is published in \textit{Research Policy}, 55 (2026), Article 105568.\\
\url{https://doi.org/10.1016/j.respol.2026.105568}
\end{center}
\vspace{0.4em}
\begin{abstract}

Generative AI can make early research work easier to produce and harder to interpret. This paper develops a compact game-theoretic model of this production--evaluation tension in the pre-doctoral academic labor market. In the model, PIs organize RA labor, allocate AI between routine and novel tasks, and choose mentoring intensity. RAs choose effort, while admissions committees infer research potential from noisy task-level signals under fixed admissions capacity. A mechanism-preserving simulation examines whether the model's qualitative mechanisms continue to hold when RAs and PIs are heterogeneous, research outcomes partly depend on luck, admissions evaluation is noisy, and elite Ph.D. capacity is fixed.
The analysis yields three implications. First, routine-task AI can increase observable routine output while reducing the diagnostic precision of routine evidence. Second, heterogeneous PI objectives and task complementarity can lead laboratories to adopt different AI strategies, with some emphasizing scalable routine production and others emphasizing mentoring and novel-task augmentation. Third, when elite Ph.D. capacity is fixed, broad improvements in visible records can raise admissions cutoffs rather than expand access proportionally. The simulation reinforces these mechanisms by showing that AI can raise routine output while weakening the link between evaluated scores and latent ability, increasing the risk that high-ability or high-realized-merit candidates are missed. The paper suggests that as routine evidence loses diagnostic content, evaluation should place greater weight on less easily automated forms of contribution, including judgment, interpretation, research design, and process-based evidence.

\vspace{0.3cm}
\noindent
\textit{\textbf{Keywords: }%
Generative AI; Pre-doctoral labor market; Research evaluation; Signal extraction; Task-based production; Academic tournaments; AI augmentation}
\noindent
\end{abstract}
}

\section{Introduction}

Generative AI creates a production--screening tension in early-career research markets. It can make research output easier to produce and harder to interpret. A cleaned dataset, polished table, well-documented script, or clear literature summary may still help a project, but it may reveal less about the researcher's own effort, judgment, and potential. This matters because early-career research output is also a screening device: it helps supervisors, admissions committees, and hiring institutions decide who has research promise. Evidence that generative AI can raise productivity while compressing performance differences or reducing output diversity makes this concern especially salient \citep{noy2023experimental,doshi2024generative,dellacqua2026jagged}. This paper studies this production--screening tension in the pre-doctoral academic labor market.

The setting is important because pre-doctoral work has become an increasingly important gateway into elite doctoral training in economics and related quantitative social sciences \citep{stansbury2023economics, stansbury2022socioeconomic}. Admissions committees allocate scarce training slots under uncertainty about future research ability, while Principal Investigators (PIs) need skilled research labor for data work, coding, literature review, and project development \citep{conley2014research, card2013nine, heckman2020publishing}. Following \citet{mangematin2000phd}'s view of doctoral training as both career investment for students and research labor for supervisors, we view pre-doctoral appointments as serving three roles at once: labor input, apprenticeship, and extended signal-generation stage.\footnote{Mangematin studies the PhD stage rather than the U.S.-style pre-doctoral stage. We use the analogy only to motivate the broader research-training logic: early research positions can combine current research labor, career investment, and later evaluation.}

Before generative AI, this market already relied on a fragile informational arrangement. Research Assistants (RAs) accepted temporary positions for wages, experience, training, and recommendation capital, while PIs obtained motivated research labor and admissions committees used work records and recommendations to infer latent research potential \citep{stansbury2023economics}. Because effort, judgment, mentoring quality, and recommendation credibility are difficult to contract on fully, the PI--RA relationship is not a simple spot-market exchange \citep{baker2002relational, levin2003relational}. These signals matter at the market level because elite opportunities are scarce and candidates compete not only against an absolute standard but also against one another \citep{lazear1981rank, hopkins2012job, hopkins2023is}.

This paper develops an analytical model of how generative AI changes this arrangement. PIs take an AI capability frontier as given and choose normalized team intensity, AI allocation between routine and novel tasks, and mentoring intensity. RAs differ in ability and choose effort. Routine and novel task outputs generate noisy task-level signals, and admissions committees use a linear evaluation rule to summarize research potential from those signals. Elite PhD admissions are modeled as a fixed-capacity tournament. The model links three objects that are usually considered separately: laboratory production, evaluation from pre-doctoral output, and capacity-constrained selection.

The model yields three main findings. First, routine-task AI can raise observable routine output while reducing the diagnostic precision of routine evidence. Polished routine artifacts become more abundant, but they may reveal less about the RA's own effort and judgment because the artifact contains a larger AI-generated component that evaluators cannot perfectly separate from human contribution. Second, AI can generate segmented laboratory strategies. Quantity-oriented PIs tilt toward scalable routine production, while quality-oriented PIs place greater value on mentoring and novel-task augmentation. Third, fixed-capacity admissions can convert broad improvements in visible records into a higher cutoff rather than broader access. Together, these findings show that AI can improve research production while changing the informational meaning of pre-doctoral output and intensifying competition for scarce academic opportunities.

The analysis distinguishes productivity, learning, and signaling effects. In a bottlenecked research-training market, these effects need not move together. Some AI-enabled gains may increase real research output or human-capital formation; others may mainly raise visible records in a fixed-capacity tournament. 
The baseline model therefore isolates the production-and-evaluation mechanism, while the mechanism-preserving simulation in Section~\ref{sec:model_faithful_simulation} examines whether the same mechanisms appear when RA talent is distributed, PI orientation is continuous, true merit contains project luck, and admissions scores are noisy. The appendix treats learning, welfare, and project-risk extensions as supporting technical material.

The contemporary U.S.-style economics pre-doctoral market provides the institutional template because it is visible, formalized, and well documented \citep{stansbury2023economics, abramitzky2024climbing}. The mechanism is not meant to describe all academic labor markets. It applies most directly to settings in which prospective doctoral students work as temporary RAs, supervisors mediate evaluation through recommendations, and access to elite doctoral training is bottlenecked \citep{camara2023reputation, hopkins2012job, lazear1981rank}.

\section{Related Literature}

This section positions the paper at the intersection of four literatures that map onto the model's main margins: the pre-doctoral labor-market setting, task-based AI in scientific work, the PI--RA training relationship, and evaluation under noisy evidence and fixed capacity. Existing work has studied these margins largely as separate problems. Our model connects them by showing how generative AI changes both the production of early-career research work and the evidentiary value of that work in a bottlenecked selection market.

\subsection{Research Labor Markets, Academic Pipelines, and Capacity Constraints}

Pre-doctoral work sits between research production and doctoral selection: it provides labor to PIs, training to prospective doctoral students, and evidence for admissions committees. This setting is shaped by uncertainty about future research productivity and by bottlenecks in elite academic placement \citep{conley2014research, card2013nine, ellison2002slowdown, heckman2020publishing}. The growth of the economics pre-doctoral stage can be understood as institutional responses from Universities to these frictions, with consequences for access and sorting into the profession \citep{stansbury2023economics, stansbury2022socioeconomic}. These studies motivate our treatment of the pre-doctoral market as a competitive pipeline in which scarce elite opportunities are allocated under uncertainty.

The setting also connects to the literature on doctoral labor markets. \citet{mangematin2000phd} treats PhD training as both students' investment in future careers and supervisors' source of research labor. Our pre-doctoral setting plays an analogous role one step earlier: current research work is tied to later selection and professional trajectories. This motivates two core objects in the model: the RA's continuation value of elite PhD admission and the PI's demand for research labor.

The same literature motivates our treatment of PI heterogeneity. Concentrated rewards for elite placement and high-impact output motivate the quality-oriented side of the model \citep{card2013nine, heckman2020publishing, manso2011incentives}. Publication pressure and measurable-output incentives motivate the quantity-oriented side \citep{hamermesh2013six, edwards2017academic}. Work on scientific competition further shows that evaluation pressure can generate quality distortions or inefficient strategic behavior \citep{kapeller2016emergent, hill2025race}. We therefore model quality-oriented and quantity-oriented PIs as facing different research incentives, rather than as morally different types.

\subsection{AI in Science, Discovery, and Task-Based Research Production}

The second literature motivates the production side of the model. We treat generative AI as task-specific rather than as a uniform productivity shock. This follows the routine/nonroutine task tradition in \citet{autor2003skill} and the task-based automation framework in \citet{acemoglu2018modeling}, \citet{acemoglu2019automation}, \citet{acemoglu2020unpacking}, \citet{acemoglu2022tasks}, and \citet{acemoglu2024taskbased}. The core insight is that technologies affect activities differently: they may substitute for human labor in some tasks, complement it in others, and create new tasks. This task-based view motivates our distinction between routine and novel research work.

Research on AI in science sharpens this task-based framing. Emerging technologies are characterized not only by novelty and rapid growth, but also by uncertainty and the capacity to reorganize existing practices \citep{rotolo2015emerging}. AI has similarly been described as an emerging general method of invention that reshapes discovery processes and the organization of scientific work \citep{bianchini2022artificial}. In a formal model of scientific discovery, \citet{agrawal2024artificial} show that AI can improve prioritized search over hypothesis spaces, but that the value of better search depends on complementary testing capacity. These studies support treating AI in science as a reorganization of research practice, not as a uniform productivity shifter.

Recent generative-AI evidence motivates the model's distinction between automation, augmentation, and diagnostic compression. AI can increase speed and quality in structured writing and knowledge-work tasks, but its effects vary across tasks and workers \citep{noy2023experimental, brynjolfsson2025generative, dellacqua2026jagged}. Evidence that AI can reduce output diversity or compress observable performance differences motivates the possibility that AI-assisted output may become less informative about individual contribution \citep{doshi2024generative, noy2023experimental}. This supports the model's separation between the productive value of AI-assisted output and its evidentiary value for evaluation.

The model also treats novel research work as uncertain and partly nonroutine. 
Novel tasks may include problem formulation, research design, hypothesis generation, interpretation, and judgment. This treatment follows problem-solving models of knowledge production \citep{garicano2000hierarchies}, research on exploratory innovation \citep{manso2011incentives}, and recent work that characterizes AI in science as reshaping discovery, hypothesis search, and methods of invention \citep{bianchini2022artificial,agrawal2024artificial}.
It is also consistent with evidence that scientific value can emerge outside initially targeted categories: \citet{aslan2024unexpectedness} document substantial unexpectedness in NIH-funded medical research.\footnote{Aslan et al. study medical research funding rather than pre-doctoral labor markets. We use their finding to motivate the broader point that scientific value may emerge outside pre-specified categories, which supports the model's treatment of novel research output as uncertain and upper-tailed.} This supports modeling breakthrough-oriented output as upper-tailed and not fully captured by routine, pre-specified tasks.

Finally, the CES research technology gives the routine--novel distinction a production interpretation. The CES form represents different degrees of substitutability across productive inputs \citep{arrow1961capital}. When routine and novel outputs are relatively substitutable, routine execution can partly compensate for weaker novel input. When they are relatively complementary, routine execution is valuable only when paired with research design, interpretation, and judgment. The complementarity case is related to O-ring production logic, where complex output depends on multiple tasks being performed well together \citep{kremer1993oring}. It is also consistent with innovation and R\&D studies that use flexible production structures to examine whether research inputs are substitutes or complements \citep{ceccagnoli2014behind, growiec2023randd}. In our setting, the elasticity parameter governs whether AI-enabled routine scaling can support research production on its own, or whether its value depends on mentored novel contribution.

\subsection{Mentoring, Apprenticeship, and Research Organizations}

The third literature motivates the PI--RA training relationship. Pre-doctoral research work is not a simple spot-market labor exchange. Many relevant margins are difficult to contract on directly: RA initiative, care in execution, PI guidance, feedback quality, and the interpretation of RA performance. Relational-contract models provide useful background for treating effort, supervision, and communication as meaningful margins under incomplete contracting \citep{baker2002relational, levin2003relational, board2015relational, watson2021theoretical}. In the baseline model of this paper, this literature supports mentoring and task assignment; it does not require a full dynamic reputation model.

The apprenticeship literature further motivates mentoring as more than a symbolic input. \citet{chassang2010building} shows how routines and cooperation can be learned under incomplete contracting. \citet{fudenberg2019training} and \citet{kostadinov2022learning} model how training, effort, and future capability evolve together. This supports the paper's learning extension and the interpretation of mentoring as an input into both current performance and future human capital. It also clarifies why AI may change not only which tasks RAs perform, but also the value of supervision and guided learning.

Two related perspectives qualify the baseline model without becoming additional mechanisms. First, absorptive capacity suggests that a common technological frontier need not imply equal effective AI capability across laboratories \citep{cohen1990absorptive}.\footnote{A simple extension would write effective AI capability as \(K_i=A_iK\), where \(A_i\) captures laboratory-level absorptive capacity. The baseline holds \(K\) common to isolate the allocation and evaluation mechanisms.} Second, research productivity is embedded in collaboration structures and cross-community ties \citep{rotolo2013centrality}. These perspectives clarify why mentoring, research context, and organizational capacity may matter when routine artifacts become easier to produce with AI.

\subsection{Evaluation, Signaling, Cumulative Advantage, and Tournaments}

The fourth literature motivates the information and allocation side of the model. Classic signaling and adverse-selection models show why evaluators infer latent quality from imperfect observables \citep{akerlof1970market, spence1973job}. Because pre-doctoral work generates graded task evidence rather than a binary credential, models of graded and noisy signals are especially relevant \citep{daley2014market, heinsalu2018dynamic, bao2021signal}. This literature motivates our use of linear prediction weights based on task-level informativeness, rather than fixed exogenous weights.

Recommendation environments add credibility concerns. \citet{camara2023reputation} show how reputation and competition shape labor-market recommendations. We use this insight in reduced form through a PI-environment credibility adjustment.\footnote{The parameter \(r_\lambda\) is a reduced-form credibility adjustment, not an endogenous reputation stock. A dynamic model in which evaluators update PI credibility from later candidate performance is a natural extension but is outside the baseline.}

This evaluation problem also connects to research assessment. Responsible-assessment frameworks warn against treating easily measured outputs as direct proxies for quality \citep{hicks2015leiden, cagan2013dora}. Performance-based research funding can shape behavior through prestige and ranking incentives \citep{hicks2012performance}, and structurally diverse teams can be disadvantaged in ex ante evaluation despite stronger ex post performance \citep{banalestanol2019evaluation}. These studies support the model's concern that shifting away from routine evidence toward more complex evidence of judgment may make evaluation more difficult, not automatically better.

Fixed-capacity competition links the evaluation problem to ranking. Tournament models motivate our treatment of admissions as relative allocation under scarce positions \citep{lazear1981rank, hopkins2012job, hopkins2023is}. This links the model to broader evidence on publication bottlenecks and escalating academic standards \citep{card2013nine, ellison2002slowdown, heckman2020publishing}. Science studies also emphasizes cumulative advantage: recognition and resources can compound over time \citep{merton1968matthew, bol2018matthew}, and institutional position can affect grant allocation \citep{viner2004institutionalized}. We do not model this dynamic directly. The literature helps interpret one implication of the model: when routine evidence becomes less diagnostic, evaluators may rely more heavily on contextual cues such as PI credibility, institutional affiliation, or prior placement history.

Together, these literatures motivate the paper's mechanism: generative AI changes task production inside laboratories; task outputs become noisy evidence about early-career researchers; and scarce admissions capacity converts that evidence into competitive outcomes. What remains underdeveloped is the link among these margins. Existing work does not explain how the same AI-assisted output can be productive for research, less informative for evaluation, and strategically consequential under fixed capacity. The model provides that link.

\section{The Model}
\label{sec:model}

This section defines the primitives of the model. We focus on a stylized version of the contemporary U.S.-style pre-doctoral academic labor market, where temporary RA appointments, recommendation-based evaluation, and competition for a fixed number of elite PhD positions are especially salient \citep{stansbury2023economics}. The model captures a setting in which PIs organize research work, RAs generate observable performance, and admissions committees infer latent research potential from imperfect evidence.

In the baseline model, each PI takes the AI capability frontier \(K\geq0\) as given.\footnote{The common-frontier assumption should be read as a baseline device. It does not imply equal effective AI capability across laboratories. Differences in infrastructure, data access, training, and verification routines could be represented by laboratory-specific effective capability \(K_i=A_iK\).} The parameter \(K\) summarizes the external AI environment available for research work. A larger \(K\) means that AI systems are more capable for tasks such as search, coding, drafting, and data analysis. The baseline does not model PI-level AI adoption or investment. Instead, it studies how laboratories allocate a given AI frontier across research tasks and mentoring choices. This convention follows the task-based view that new technologies affect the productivity and feasibility of particular tasks, rather than acting only as uniform total-factor-productivity shocks \citep{autor2003skill, acemoglu2018modeling, acemoglu2019automation, agrawal2019economics, korinek2023generative}.

A PI has type \(\lambda\in\{\lambda_Q,\lambda_N\}\), where \(\lambda_Q\) denotes a quality-oriented PI and \(\lambda_N\) denotes a quantity-oriented PI. Each PI chooses
$
(n_\lambda,\alpha_\lambda,g_\lambda).
$
The first choice, \(n_\lambda\geq0\), is normalized team intensity: a continuous measure of filled RA labor operated by the PI in equilibrium, rather than a literal headcount or a stock of posted vacancies. Participation and sorting determine which RA types fill those positions; unfilled postings are outside the baseline model. The second choice, \(\alpha_\lambda\in[0,1]\), is the share of the available AI frontier allocated to routine tasks. The third choice, \(g_\lambda\geq0\), is mentoring intensity, which captures the PI's time and effort devoted to supervision, feedback, and guidance.

Given \(K\) and \(\alpha_\lambda\), routine-task AI is
\[
K^T_{R,\lambda}\equiv \alpha_\lambda K,
\]
and novel-task AI is
\[
K^T_{N,\lambda}\equiv (1-\alpha_\lambda)K.
\]
Thus,
\[
K^T_{R,\lambda}\in[0,K],
\qquad
K^T_{N,\lambda}\in[0,K],
\qquad
K^T_{R,\lambda}+K^T_{N,\lambda}=K.
\]
The superscript \(T\) denotes task allocation; the subscripts \(R\) and \(N\) refer to routine and novel tasks, respectively.\footnote{The subscript \(N\) in \(K^T_{N,\lambda}\) denotes novel tasks, while \(\lambda_N\) denotes quantity-oriented PIs.}

The model follows a pre-doctoral RA through this environment. A PI first chooses filled team intensity, AI allocation, and mentoring. The RA then enters the lab with an underlying research ability and chooses effort. Routine and novel tasks generate outputs, which enter the PI's research production and also create evidence about the RA. Admissions committees observe noisy task-level signals, evaluate research potential, and allocate a fixed mass of elite PhD slots through a capacity-constrained admissions stage.

The production side and the evaluation side are deliberately separated. On the production side, routine and novel task outputs are combined through a CES research technology, allowing the degree of substitutability between structured execution and novel judgment to differ across research environments \citep{acemoglu2018modeling, kremer1993oring}. On the evaluation side, admissions committees place weight on observed task performance according to its informativeness, rather than according to fixed exogenous signal weights. This information structure is motivated by models of noisy signals, market inference, and tournament-like allocation under scarcity \citep{daley2014market, hopkins2012job}. Learning and welfare are introduced later as extensions in the appendix. Fig.~\ref{fig:mechanism_overview} provides an overview of the mechanism, and Table~\ref{tab:notation_bayes} summarizes the notation.

\begin{figure}[t]
\centering
\begin{adjustbox}{max width=\linewidth}
\begin{tikzpicture}[
    font=\small,
    box/.style={
        draw,
        rounded corners=2pt,
        line width=0.6pt,
        fill=white,
        align=center,
        inner sep=4pt
    },
    arrow/.style={
        -{Latex[length=2mm]},
        line width=0.6pt
    },
    node distance=8mm and 10mm
]

\node[box, text width=34mm] (inputs)
{AI capability \(K\)\\
PI choices \((n,\alpha,g)\)\\
RA ability \(\theta\)};

\node[box, text width=33mm, right=18mm of inputs] (prod)
{Task production\\[2pt]
RA effort \(e\)\\
Routine output \(y_R\)\\
Novel output \(y_N\)};

\node[box, text width=31mm, right=14mm of prod, yshift=9mm] (ces)
{Research output \(Y_\lambda\)\\[2pt]
(CES aggregator)};

\node[box, text width=31mm, right=14mm of prod, yshift=-9mm] (signals)
{Observed signals\\[2pt]
\(s_R=a_R+\rho_R h_R+\varepsilon_R\)\\
\(s_N=y_N+\varepsilon_N\)};

\node[box, text width=31mm, right=14mm of signals] (score)
{Evaluation score\\[2pt]
\(\hat{\theta}\) and \(S=r_\lambda\hat{\theta}\)};

\node[box, text width=35mm, right=14mm of score] (admit)
{Fixed-capacity tournament\\[2pt]
Cutoff \(S^*\) and slots \(Q\)\\
Admission probability \(P_\lambda\)};

\path (inputs.north) coordinate (stage1);
\path ($(prod.north)!0.5!(ces.north)$) coordinate (stage2);
\path ($(score.north)!0.5!(admit.north)$) coordinate (stage3);

\node[font=\small\bfseries, above=7mm of stage1] {Stage 1: Organization};
\node[font=\small\bfseries, above=7mm of stage2] {Stage 2: Production and Signals};
\node[font=\small\bfseries, above=7mm of stage3] {Stage 3: Evaluation and Admission};

\draw[arrow] (inputs.east) -- (prod.west);
\draw[arrow] (prod.east) -- ++(5mm,0) |- (ces.west);
\draw[arrow] (prod.east) -- ++(5mm,0) |- (signals.west);
\draw[arrow] (signals.east) -- (score.west);
\draw[arrow] (score.east) -- (admit.west);

\end{tikzpicture}
\end{adjustbox}
\caption{Mechanism overview. Stage 1 combines the exogenous AI capability frontier \(K\), PI organizational choices, and RA heterogeneity. Stage 2 separates into two branches. On the production side, RA effort and task outputs generate research output \(Y_\lambda\) through the CES aggregator. On the evaluation side, the market observes only noisy signals \((s_R,s_N)\). Stage 3 converts the resulting evaluation score into admissions outcomes through a fixed-capacity cutoff \(S^*\). The distinction between the production branch and the evaluation branch is a central feature of the model. This is a schematic overview of model objects, not a calibrated quantitative result.}
\label{fig:mechanism_overview}
\end{figure}

\subsection{Players}
\label{sec:players}

The model consists of three types of players.

\begin{itemize}

\item \textbf{Principal Investigators (PIs).} The population of PIs is normalized to have total mass one. PIs organize research projects, supervise RAs, provide mentoring, and write recommendations. These roles involve non-contractible margins such as feedback quality, task assignment, and the interpretation of RA performance \citep{baker2002relational, levin2003relational}. In the baseline model, these margins motivate the PI's choices over team organization, AI allocation, and mentoring.

PIs differ in their research objectives. A fraction \(\mu\in(0,1)\) are quality-oriented, denoted by type \(\lambda_Q\), and place greater weight on high-value or breakthrough research outcomes. The remaining fraction \(1-\mu\) are quantity-oriented, denoted by type \(\lambda_N\), and place greater weight on scalable research throughput. This distinction is a reduced-form representation of different research incentives, not a moral classification. For compact notation, let \(\mu_{\lambda_Q}\equiv\mu\) and \(\mu_{\lambda_N}\equiv1-\mu\), so that \(\mu_\lambda\) denotes the mass of PIs of type \(\lambda\).

\item \textbf{Research Assistants (RAs).} RAs enter the model as prospective doctoral students working in temporary pre-doctoral research positions. The pre-doctoral stage is modeled as a transitional period before PhD admissions: it provides labor input to the PI's research projects, but it also creates training opportunities and evidence about the RA's research potential. This treatment follows a broader view of early research training as a transitional labor-market stage in which current research work is tied to later professional trajectories \citep{mangematin2000phd}. It is also consistent with human-capital theory and apprenticeship-style models in which effort, supervision, and learning evolve together during early-career training \citep{becker1962investment, fudenberg2019training, kostadinov2022learning}.

Each RA has underlying research ability
\[
\theta \sim \mathrm{TN}(\mu_\theta,\sigma_\theta^2;\theta_L,\theta_H),
\qquad 0<\theta_L<\theta_H<\infty,
\]
which is private information. Here \(\mathrm{TN}(\cdot)\) denotes a normal distribution truncated to the compact ability support \([\theta_L,\theta_H]\).\footnote{The compact support is a production-domain regularity condition. It keeps task outputs and effort-capacity terms well defined on the relevant equilibrium domain.}
During the pre-doctoral period, the RA chooses continuous effort \(e\in[0,\infty)\). Ability is not directly observed by the admissions market, so it must be inferred from subsequent task-level performance and other evidence generated during the appointment \citep{conley2014research, spence1973job, daley2014market}.

\item \textbf{The Market.} The ``Market'' represents the collective of top-tier PhD admissions committees. It does not observe \(\theta\) directly. Instead, it observes noisy task-level signals generated in the PI's lab and evaluates the candidate's research potential from those signals. This information structure follows signaling models and models of graded noisy evidence \citep{spence1973job, daley2014market, heinsalu2018dynamic}. Because admission is to a fixed number of elite PhD positions, the market also has the structure of a rank-order tournament: relative standing matters in addition to absolute performance \citep{lazear1981rank, hopkins2012job}. The admissions stage therefore combines signal extraction with relative ranking under scarcity.

\end{itemize}

\subsection{Research Production, Observed Signals, and AI}
\label{sec:production_signals}

We model research as a set of tasks rather than as a single undifferentiated output. This choice reflects the task-based view of technological change: new technologies affect different activities in different ways, rather than simply raising productivity everywhere by the same amount \citep{autor2003skill, acemoglu2018modeling, acemoglu2019automation, acemoglu2020unpacking, acemoglu2024taskbased}. We distinguish between two kinds of research work:
\begin{itemize}
    \item \textbf{Routine tasks (\(T_R\)):} structured tasks such as literature search, data cleaning, standardized regressions, code debugging, formatting, and other forms of execution.
    \item \textbf{Novel tasks (\(T_N\)):} less structured tasks such as hypothesis generation, research design, causal reasoning, problem formulation, and interpretation of ambiguous findings.
\end{itemize}


Define human input for the production function as
\[
h_R\equiv \theta+\eta_R e,
\qquad
h_N\equiv \theta+\eta_N e+\psi g,
\]
where \(\eta_R,\eta_N,\psi>0\). The term \(h_R\) is the human input into routine tasks, while \(h_N\) is the mentored human input into novel tasks. The term \(\psi g\) captures the incremental contribution of PI mentoring to novel-task human input, where \(g\) is mentoring intensity and \(\psi>0\) measures how effectively mentoring translates into research design, interpretation, and other judgment-intensive contributions.\footnote{This reduced-form specification follows human-capital and apprenticeship models in which ability, effort, and training jointly shape current performance and future capability \citep{becker1962investment,fudenberg2019training,kostadinov2022learning}. The distinction between routine and novel human inputs is also consistent with the task-based view that different activities rely on different combinations of human input and technology \citep{autor2003skill,acemoglu2018modeling,acemoglu2019automation}. }
Both \(h_R\) and \(h_N\) are reduced-form human-input terms rather than separately observed objects.

Following the task-based framework pioneered by \citet{autor2003skill,acemoglu_nd_skills} and \citet{acemoglu2022tasks}, we model the output of routine tasks as a linear combination of AI's direct automation value and human contribution:
\begin{equation}
y_R
=
a_R(K_R^T)+m_R(K_R^T)h_R.
\label{eq:yR_bayes}
\end{equation}
The first term, \(a_R(K_R^T)\geq0\), is the direct automation value of routine-task AI, with \(a_R'(\cdot)>0\). The second term, \(m_R(K_R^T)h_R\), is the human contribution to routine output. We assume \(m_R(K_R^T)>0\) and \(m_R'(\cdot)\geq0\), so routine-task AI weakly increases the productive contribution of human routine input.
The key modeling distinction in the paper is that productive routine output and evidentiary value need not move together.

The admissions market does not observe the RA's human input \(h_R\) directly. We summarize routine-task evidence by the reduced-form signal
\begin{equation}
s_R
=
a_R(K_R^T)+\rho_R(K_R^T)h_R+\varepsilon_R,
\qquad
\varepsilon_R\sim N(0,\sigma_R^2).
\label{eq:sR_bayes}
\end{equation}
Here \(a_R(K_R^T)\) is the observable automation component of routine-task output. It can raise the surface quality of routine artifacts, such as cleaned data, code, tables, or written summaries, but it is not itself diagnostic of the RA's own ability or effort. The term \(\rho_R(K_R^T)>0\) is the diagnostic loading of routine evidence on the RA's human input.\footnote{
A simple artifact-level microfoundation delivers a decreasing diagnostic loading. Suppose that, after conditioning on task allocation and deterministic routine-production scale, the evaluator observes a normalized routine artifact
$
z_R
=
h_R+A_R(K_R^T)+\nu_R,
\qquad
A_R(K_R^T)\sim N\!\left(0,\sigma_A^2(K_R^T)\right),
\qquad
\nu_R\sim N(0,\sigma_\nu^2),
$
where \(A_R(K_R^T)\) is an inseparable AI-generated component and \(\nu_R\) is ordinary artifact noise. In the local residualized benchmark, \(A_R(K_R^T)\), \(\nu_R\), and the residual innovation in \(h_R\) are mutually independent. This representation follows the standard signal-extraction logic in which evaluators infer latent quality from noisy evidence rather than directly observing ability, effort, or contribution \citep{spence1973job, daley2014market, heinsalu2018dynamic, bao2021signal}. Let \(\sigma_h^2>0\) denote the local residual variance of \(h_R\). If routine-task AI increases the variance of the inseparable AI-generated component,
$
\frac{d\sigma_A^2(K_R^T)}{dK_R^T}>0,
$
then the best linear predictor of \(h_R\) from \(z_R\) places weight
$
\rho_R(K_R^T)
=
\frac{\sigma_h^2}
{\sigma_h^2+\sigma_A^2(K_R^T)+\sigma_\nu^2}.
$
Differentiating gives
$
\frac{d\rho_R(K_R^T)}{dK_R^T}
=
-
\frac{
\sigma_h^2\,d\sigma_A^2(K_R^T)/dK_R^T
}
{
\left[\sigma_h^2+\sigma_A^2(K_R^T)+\sigma_\nu^2\right]^2
}
<0.
$
Thus, \(\rho_R'(K_R^T)<0\) follows from the attribution-risk primitive that routine-task AI raises the variance of the inseparable non-human component of the observed artifact. 
}
This loading differs from \(m_R(K_R^T)\) in the production function: \(m_R(\cdot)\) governs the productive contribution of human input to routine output, whereas \(\rho_R(\cdot)\) governs the evidentiary value of routine output for evaluation. The noise term \(\varepsilon_R\) captures residual measurement and attribution uncertainty in the reduced-form signal. In the baseline, \(\sigma_R^2\) is held fixed to isolate the loading channel; if residual signal noise also varied with \(K_R^T\), the corresponding precision term would be \(\rho_R(K_R^T)^2/\sigma_R^2(K_R^T)\).

This signal structure follows models in which evaluators infer latent quality from imperfect or graded evidence rather than observing ability and effort directly \citep{spence1973job,daley2014market,heinsalu2018dynamic,bao2021signal}. It also makes precise why a polished table, clean code, or well-written summary may remain useful for the project while becoming less informative for evaluation: the routine artifact contains more AI-generated variation that cannot be perfectly attributed to the RA.\footnote{The microfoundation is motivated by evidence that generative AI can raise output quality while compressing observable performance differences, reducing output diversity, or making individual contribution harder to separate from tool assistance in structured tasks \citep{noy2023experimental,doshi2024generative,dellacqua2026jagged,brynjolfsson2025generative}. It is not a claim that AI always reduces informativeness. In some settings, AI may improve documentation, traceability, or auditability; such process evidence would reduce the effective \(\sigma_A^2(K_R^T)\) or add an independent signal.}

Similarly, novel task output is
\begin{equation}
y_N
=
\kappa_\lambda(K_N^T)h_N,
\label{eq:yN_bayes}
\end{equation}
where \(\kappa_\lambda(0)=1\) and \(\kappa_\lambda'(K_N^T)>0\). The function \(\kappa_\lambda\) captures AI augmentation in novel tasks. AI may help with ideation, coding support, feedback, interpretation, and problem formulation, but its value still depends on human judgment and mentoring. This view is consistent with problem-solving models of knowledge production, exploratory innovation, and recent work on generative AI in research and human--AI collaboration \citep{garicano2000hierarchies, manso2011incentives, korinek2023generative, fugener2025roles}.\footnote{This treatment is also consistent with evidence that scientific projects can generate outputs outside initially targeted categories \citep{aslan2024unexpectedness}, which motivates the uncertain and upper-tailed interpretation of novel research output. The mechanism-preserving simulation uses the linear approximation \(\kappa_\lambda(K_N^T)=1+\phi_\lambda K_N^T\). The theoretical results do not depend on this linear form. We maintain \(y_R>0\) and \(y_N>0\) on the equilibrium support so that the CES aggregator is well defined.}

The market observes the routine signal in \eqref{eq:sR_bayes} and the novel signal
\begin{equation}
s_N
=
y_N+\varepsilon_N,
\qquad
\varepsilon_N\sim N(0,\sigma_N^2).
\label{eq:sN_bayes}
\end{equation}
The noise terms capture luck, measurement error, and evaluative uncertainty. This follows stochastic production and noisy-signal models in which evaluators infer quality from imperfect observables \citep{just1978stochastic, daley2014market, heinsalu2018dynamic}.

To capture how routine execution and novel judgment combine in research production, we use a CES aggregator:
\begin{equation}
Y_\lambda
=
\left[
\tau y_R^{\frac{\varsigma_\lambda-1}{\varsigma_\lambda}}
+
(1-\tau)y_N^{\frac{\varsigma_\lambda-1}{\varsigma_\lambda}}
\right]^{\frac{\varsigma_\lambda}{\varsigma_\lambda-1}},
\qquad
\tau\in(0,1),\ \varsigma_\lambda>0.
\label{eq:CES}
\end{equation}
The CES form represents different degrees of substitutability across productive inputs \citep{arrow1961capital}. When \(\varsigma_\lambda=1\), \eqref{eq:CES} is interpreted by its Cobb--Douglas limit. When \(\varsigma_\lambda>1\), routine and novel outputs are relatively substitutable: additional routine execution can partly compensate for weaker novel input. When \(\varsigma_\lambda<1\), routine and novel outputs are relatively complementary: routine execution is valuable, but it cannot easily replace research design, interpretation, or original judgment. The complementarity case is related to O-ring production logic and to models of AI-assisted scientific discovery in which better hypothesis generation requires complementary testing and interpretation capacity \citep{kremer1993oring, agrawal2024artificial}. Innovation and R\&D studies also use CES or flexible production structures to study whether research inputs are substitutes or complements \citep{ceccagnoli2014behind, growiec2023randd}.\footnote{The elasticity \(\varsigma_\lambda\) should be read as a reduced-form feature of the research environment rather than as a pure PI preference parameter. PI objectives may be correlated with research environments, but the model does not require preferences and physical task technology to be identical.}

The section therefore separates two objects that are often conflated. Task outputs \((y_R,y_N)\) enter the PI's research production technology. Observed signals \((s_R,s_N)\) enter the admissions market's evaluation problem. This distinction allows the model to capture both the productive value of AI-assisted work and the changing evidentiary value of that work for candidate evaluation \citep{acemoglu2018modeling, daley2014market, bao2021signal}.

\subsection{Evaluation from Noisy Signals}
\label{sec:information_structure}

The admissions market does not observe an RA's research ability directly. Ability \(\theta\) is private information, effort \(e\) is not directly contractible, and admissions committees observe only noisy evidence generated by the RA's work. This places the model in a standard setting of evaluation under imperfect information: evaluators infer latent quality from observed but noisy signals \citep{akerlof1970market, spence1973job, daley2014market, heinsalu2018dynamic, bao2021signal}.

To keep the inference problem tractable, we use a linear-Gaussian benchmark. With a Gaussian prior and Gaussian noise, the posterior mean is a linear, precision-weighted average of the prior mean and observed signals \citep{degroot1970optimal, gelman2013bayesian}. Because ability has bounded support in the main model, the linear formula is exact only in the corresponding unbounded Gaussian benchmark. Under bounded support, we interpret it as a best linear prediction rule. The online appendix derives the exact truncated-normal posterior and shows that the main informativeness result is unchanged.\footnote{Truncation adds a nonlinear boundary correction to the posterior mean. This correction matters near the lower and upper support boundaries, but it does not change the limiting result that routine evidence ceases to affect the posterior as \(\rho_R(K_R^T)\to0\).}

The market is assumed to condition on the PI environment, the prevailing AI frontier, and the task-level AI allocation when forming its evaluation.\footnote{This is a benchmark information assumption. If evaluators are uncertain about the laboratory environment or AI allocation, that uncertainty can be represented as additional measurement noise in the task-level signals.} Define the centered routine signal as
\begin{align}
\tilde s_R
&\equiv
s_R-a_R(K_R^T)-\rho_R(K_R^T)\mu_\theta
\nonumber\\
&=
\rho_R(K_R^T)\bigl(\theta-\mu_\theta+\eta_R e\bigr)+\varepsilon_R.
\label{eq:yR_tilde}
\end{align}
Similarly, define the centered novel signal as
\begin{align}
\tilde s_N
&\equiv
s_N-\kappa_\lambda(K_N^T)\mu_\theta
\nonumber\\
&=
\kappa_\lambda(K_N^T)\bigl(\theta-\mu_\theta+\eta_N e+\psi g\bigr)+\varepsilon_N.
\label{eq:yN_tilde}
\end{align}
Neither centered signal is a pure measure of ability. Both combine ability, effort, mentoring, and noise. The weights below should therefore be read as benchmark linear prediction weights after deterministic effort and mentoring components are residualized, or equivalently as the market's best linear evaluation rule in the maintained linearized environment.\footnote{The online appendix defines the residualized benchmark environment used for the technical derivation, without assuming that evaluators directly observe effort.}

The signal noises are assumed to be independent:
\[
\varepsilon_R\sim N(0,\sigma_R^2),
\qquad
\varepsilon_N\sim N(0,\sigma_N^2).
\]
Under the linear-Gaussian benchmark, the market's summary evaluation is
\begin{equation}
\hat{\theta}
=
\mu_\theta
+
\omega_R^{B}(\alpha,K,\lambda)\tilde s_R
+
\omega_N^{B}(\alpha,K,\lambda)\tilde s_N.
\label{eq:theta_hat}
\end{equation}
Here \(\hat{\theta}\) is an admissions evaluation of research potential, written in ability units. It is the benchmark linear prediction rule used in the main text.

The coefficient on the routine signal is
\begin{align}
\omega_R^{B}(\alpha,K,\lambda)
&=
\frac{\rho_R(K_R^T)/\sigma_R^2}
{
\frac{1}{\sigma_\theta^2}
+
\frac{\rho_R(K_R^T)^2}{\sigma_R^2}
+
\frac{\kappa_\lambda(K_N^T)^2}{\sigma_N^2}
},
\label{eq:omegaR_B}
\end{align}
and the coefficient on the novel signal is
\begin{align}
\omega_N^{B}(\alpha,K,\lambda)
&=
\frac{\kappa_\lambda(K_N^T)/\sigma_N^2}
{
\frac{1}{\sigma_\theta^2}
+
\frac{\rho_R(K_R^T)^2}{\sigma_R^2}
+
\frac{\kappa_\lambda(K_N^T)^2}{\sigma_N^2}
}.
\label{eq:omegaN_B}
\end{align}
These coefficients follow from the informativeness of the two signal channels. Routine evidence contributes precision \(\rho_R(K_R^T)^2/\sigma_R^2\), while novel evidence contributes precision \(\kappa_\lambda(K_N^T)^2/\sigma_N^2\). A signal is more informative when its loading on human contribution is larger and its noise variance is smaller. The coefficients in \eqref{eq:omegaR_B}--\eqref{eq:omegaN_B} are the corresponding linear prediction coefficients, adjusted for signal scale and prior uncertainty \citep{daley2014market, heinsalu2018dynamic, bao2021signal}.

The key implication is simple. If routine-task AI makes routine evidence less diagnostic of the RA's own contribution, then routine evidence becomes less useful for evaluation. In the limiting case,
\[
\lim_{\rho_R(K_R^T)\to0}\omega_R^B(\alpha,K,\lambda)=0.
\]
Routine output may still be valuable for producing research, but it becomes less useful for evaluating the RA.

The admissions-relevant score scales the market's evaluation by the PI-environment credibility adjustment:
\begin{equation}
S=r_\lambda\hat{\theta}.
\label{eq:S_bayes}
\end{equation}

This term captures the idea that evaluators may interpret otherwise similar task-level evidence differently depending on the context in which it was produced \citep{camara2023reputation}.\footnote{The main signal-weight mechanism does not require variation in \(r_\lambda\). Setting \(r_\lambda=1\) leaves the production-evaluation mechanism unchanged. The parameter is a reduced-form credibility adjustment, not an endogenous reputation stock.}

The model focuses on pre-doctoral research evidence because that is the channel most directly affected by AI-assisted research work. Other admissions evidence, such as coursework, grades, standardized preparation, or additional letters, could be represented as additional noisy signals in the evaluation rule. Such signals would add independent precision and may dampen the effect of diagnostic compression in routine research evidence. They do not eliminate the mechanism whenever pre-doctoral research output remains an important signal of research potential \citep{spence1973job, daley2014market}.

\subsection{Payoffs}
\label{sec:payoffs}

The RA payoff captures the trade-off between wages, effort costs, and the value of admission to an elite PhD program. The PI payoff captures the value of research output, the cost of organizing and mentoring RAs, and the difference between quantity-oriented and quality-oriented research environments.

\paragraph{RA payoffs.}

Pre-doctoral wages are treated as fixed in the short run.\footnote{This is a short-run institutional simplification. Many pre-doctoral positions are posted with standardized salary bands, and the paper focuses on effort, task allocation, and evaluation rather than wage bargaining.} Let
\[
w=\bar w
\]
denote the wage paid to an RA. We also let \(\bar U\) denote the RA's outside option, which enters the participation constraint below. Changes in compensation and outside opportunities can therefore be represented by shifts in \(\bar w\) and \(\bar U\), respectively. A higher \(\bar w\) raises the RA's payoff but also increases the PI's labor cost, while a higher \(\bar U\) tightens the participation constraint. These shifts can affect participation and organizational scale, but they are not the source of the production--evaluation mechanism studied here.

Following the quadratic effort-cost specification used in multitask and relational-contract models in\citet{holmstrom1991multitask, baker2002relational}, the RA incurs a convex effort cost:
\begin{equation}
c_\lambda(e;\theta,K^T_{R,\lambda})
=
\frac{e^2}{2\chi(\theta,K^T_{R,\lambda})},
\label{eq:cost_bayes}
\end{equation}
where \(K^T_{R,\lambda}=\alpha_\lambda K\) is the amount of AI capability allocated to routine tasks in a type-\(\lambda\) laboratory. The function \(\chi(\theta,K^T_{R,\lambda})>0\) is an effective effort-capacity term. We assume
\[
\frac{\partial \chi(\theta,K^T_{R,\lambda})}{\partial \theta}>0,
\qquad
\frac{\partial \chi(\theta,K^T_{R,\lambda})}{\partial K^T_{R,\lambda}}>0.
\]
Thus, higher-ability RAs face lower effective effort costs, and routine-task AI can make effort more productive in structured research work. This cost channel is separate from the signal-informativeness channel: routine-task AI may make effort easier while also making routine artifacts less diagnostic.

Using the evaluation rule in Section~\ref{sec:information_structure}, the admissions-relevant score can be written in reduced form as
\begin{equation}
S
=
A_\lambda^{B}(\theta;K,\alpha_\lambda,g_\lambda)
+
B_\lambda^{B}(K,\alpha_\lambda,\lambda)e
+
\Sigma_\lambda^{B}(K,\alpha_\lambda,\lambda)\xi,
\qquad
\xi\sim N(0,1).
\label{eq:S_reduced}
\end{equation}
Here \(A_\lambda^{B}\) is the deterministic part of the score, \(B_\lambda^{B}\) is the marginal loading of RA effort into the admissions score, and \(\Sigma_\lambda^{B}>0\) is residual score noise. Closed-form expressions are reported in the online appendix.

Given an admissions cutoff \(S^*\), the admission probability of an RA with ability \(\theta\) and effort \(e\) in a type-\(\lambda\) laboratory is
\begin{equation}
P_\lambda(\theta,e;S^*)
=
1-\Phi\!\left(
\frac{
S^*
-
A_\lambda^{B}(\theta;K,\alpha_\lambda,g_\lambda)
-
B_\lambda^{B}(K,\alpha_\lambda,\lambda)e
}{
\Sigma_\lambda^{B}(K,\alpha_\lambda,\lambda)
}
\right).
\label{eq:Padm_bayes}
\end{equation}

The RA's utility is
\begin{equation}
U_{RA,\lambda}(\theta,e;S^*)
=
\bar w
-
\frac{e^2}{2\chi(\theta,K^T_{R,\lambda})}
+
\beta_{RA}V P_\lambda(\theta,e;S^*),
\label{eq:URA_bayes}
\end{equation}
where \(\beta_{RA}\in(0,1]\) is the RA's discount factor and \(V>0\) is the continuation value of elite PhD admission. The continuation value \(V\) captures the career value of successful placement, consistent with research on early research training and professional trajectories \citep{mangematin2000phd}. The RA chooses effort according to
\begin{equation}
e_\lambda^*(\theta)
\in
\arg\max_{e\geq0}
U_{RA,\lambda}(\theta,e;S^*).
\label{eq:e_star}
\end{equation}
The online appendix gives sufficient regularity conditions under which this problem has a unique optimal solution in the maintained region. The solution may be interior or at the boundary; outside the maintained region, the effort choice should be interpreted as an optimal-effort correspondence.

Let
\begin{equation}
\widetilde U_\lambda(\theta)
\equiv
\max_{e\geq0}
U_{RA,\lambda}(\theta,e;S^*)
\label{eq:indirect_utility}
\end{equation}
denote the RA's indirect utility in a type-\(\lambda\) environment. The RA participates only if
\begin{equation}
\widetilde U_\lambda(\theta)\geq \bar U,
\label{eq:PC_bayes}
\end{equation}
where \(\bar U\) is the outside option.

When RAs can choose across PI segments, equilibrium assignment is defined by
\begin{equation}
\Theta_\lambda
=
\Bigl\{
\theta\in[\theta_L,\theta_H]:
\widetilde U_\lambda(\theta)
\ge
\widetilde U_{\lambda'}(\theta)
\text{ for all } \lambda'\neq\lambda,
\quad
\widetilde U_\lambda(\theta)\ge \bar U
\Bigr\}.
\label{eq:sorting_set}
\end{equation}
Thus, \(\Theta_\lambda\) is the set of abilities that choose the type-\(\lambda\) segment and participate.\footnote{The model abstracts from a separate RA-position matching market. The set \(\Theta_\lambda\) should be read as the equilibrium ability composition of RAs hired into type-\(\lambda\) positions. Without monotonicity or single crossing, \(\Theta_\lambda\) need not be an interval. The cutoff representation \(\underline{\theta}_\lambda\) is used only in the maintained region described in the appendix.} Under this convention, \(\Theta_\lambda\) determines the type composition of filled positions, while \(n_\lambda\) records their normalized mass. A model of posted positions would require a separate fill rate; the baseline abstracts from that layer. Under the monotonicity conditions stated in the online appendix, this set can be summarized by a lower cutoff \(\underline{\theta}_\lambda\) within the relevant segment.

\paragraph{PI payoffs.}

We model the PI's organizational costs as a convex function of team size and mentoring intensity. The quadratic cost of supervision and coordination captures the standard limits to the span of control driven by coordination frictions and communication costs within organizations, as formalized by \citet{garicano2000hierarchies,jones2021rise,fudenberg2019training}. A PI of type \(\lambda\) chooses normalized team intensity \(n_\lambda\geq0\), AI allocation \(\alpha_\lambda\in[0,1]\), and mentoring intensity \(g_\lambda\geq0\). In both PI types, organizational costs are
\begin{equation}
C_\lambda(n_\lambda,g_\lambda)
=
\bar w n_\lambda
+
\frac{c_g}{2}g_\lambda^2 n_\lambda
+
\frac{c_n}{2}n_\lambda^2,
\qquad
c_g,c_n>0.
\label{eq:PIcost_bayes}
\end{equation}
The first term is the wage bill. The second term captures the cost of providing intensive mentoring to a larger RA team. The third term captures supervision and coordination congestion.

Given equilibrium effort \(e_\lambda^*(\theta)\), task outputs for an RA of ability \(\theta\) are
\begin{align}
y_{R,\lambda}(\theta)
&=
a_R(K^T_{R,\lambda})
+
m_R(K^T_{R,\lambda})
\bigl(\theta+\eta_R e_\lambda^*(\theta)\bigr),
\label{eq:yR_individual_payoff}
\\
y_{N,\lambda}(\theta)
&=
\kappa_\lambda(K^T_{N,\lambda})
\bigl(\theta+\eta_N e_\lambda^*(\theta)+\psi g_\lambda\bigr).
\label{eq:yN_individual_payoff}
\end{align}

Let \(F_\lambda\) denote the conditional equilibrium distribution of abilities among RAs filling positions in the type-\(\lambda\) segment, normalized on \(\Theta_\lambda\). The per-RA expected CES research output is
\begin{equation}
\bar Y_\lambda
=
\int_{\Theta_\lambda}
\left[
\tau y_{R,\lambda}(\theta)^{\frac{\varsigma_\lambda-1}{\varsigma_\lambda}}
+
(1-\tau)y_{N,\lambda}(\theta)^{\frac{\varsigma_\lambda-1}{\varsigma_\lambda}}
\right]^{\frac{\varsigma_\lambda}{\varsigma_\lambda-1}}
dF_\lambda(\theta).
\label{eq:Ybar_bayes}
\end{equation}
This definition takes the expectation of the CES output itself, rather than applying the CES aggregator to average task outputs.

Quality-oriented research also has an upper-tail component. Let realized novel-project merit be
\[
q_N=y_{N,\lambda}(\theta)+\zeta,
\qquad
\zeta\sim N(0,\sigma_\zeta^2),
\]
where \(\zeta\) captures project-level uncertainty that is distinct from the signal noises. A breakthrough occurs when \(q_N>\bar q\). The average breakthrough probability in the type-\(\lambda\) segment is
\begin{equation}
\bar p_{B,\lambda}
=
\int_{\Theta_\lambda}
\left[
1-
\Phi\!\left(
\frac{
\bar q
-
y_{N,\lambda}(\theta)
}{
\sigma_\zeta
}
\right)
\right]
dF_\lambda(\theta).
\label{eq:pBbar_bayes}
\end{equation}
With \(n_\lambda\) independent project attempts and average success probability \(\bar p_{B,\lambda}\), the probability of at least one breakthrough is
\[
1-(1-\bar p_{B,\lambda})^{n_\lambda}.
\]
Because \(n_\lambda\) is normalized team intensity rather than literal headcount, this expression should be read as a continuous approximation to a portfolio success probability \citep{aghion2008academic}. The threshold formulation captures the idea that exploratory research has an uncertain upper tail \citep{manso2011incentives}.

Our distinction between quantity-oriented and quality-oriented PIs builds on the fundamental trade-off between exploitation and exploration in the economics of science \citep{manso2011incentives}. 

A quantity-oriented PI values scalable research output and solves
\begin{equation}
\Pi_N
=
\gamma_N n_N\bar Y_N
-
C_N(n_N,g_N),
\qquad
\gamma_N>0.
\label{eq:PIN_bayes}
\end{equation}
The term \(n_N\bar Y_N\) captures scalable research throughput. The quantity-oriented objective in equation \eqref{eq:PIN_bayes} reflects the realities of the modern academic reward system, where hypercompetition and quantitative performance metrics often incentivize scholars to prioritize scalable, safe research output \citep{edwards2017academic, azoulay2011incentives}.

A quality-oriented PI places primary value on breakthrough probability \citep{aghion2008academic}:
\begin{equation}
\Pi_Q
=
\Omega
\Bigl[
1-(1-\bar p_{B,Q})^{n_Q}
\Bigr]
+
\gamma_Q n_Q\bar Y_Q
-
C_Q(n_Q,g_Q),
\qquad
\Omega>0,\quad \gamma_Q\geq0.
\label{eq:PIQ_bayes}
\end{equation}
The first term captures the value of at least one high-impact research outcome. The second term allows quality-oriented PIs to value ordinary CES research output as well.\footnote{Because \(\bar Y_Q\) is defined as expected per-RA CES output, the ordinary-output component is written as \(n_Q\bar Y_Q\). If \(\bar Y_Q\) were defined as total portfolio output, the explicit \(n_Q\) factor would not be needed.} 

The two PI types differ in objectives and may operate in different research environments. Quantity-oriented PIs place greater weight on scalable throughput. Quality-oriented PIs place greater weight on upper-tail research success. In addition, the same AI frontier can have different effects depending on whether routine and novel tasks are more substitutable or more complementary in the laboratory's CES technology \citep{garicano2000hierarchies, kremer1993oring, growiec2023randd}.

\begin{table}[t]
\centering
\caption{Core Notation}
\label{tab:notation_bayes}
\scriptsize
\begin{tabular}{>{\raggedright\arraybackslash}p{2.4cm}>{\raggedright\arraybackslash}p{7.0cm}>{\raggedright\arraybackslash}p{3.7cm}}
\toprule
\textbf{Symbol} & \textbf{Definition} & \textbf{Domain} \\
\midrule
\multicolumn{3}{l}{\textit{Agents and types}} \\
\(\lambda\) & PI type & \(\{\lambda_Q,\lambda_N\}\) \\
\(\lambda_Q\) & Quality-oriented PI type & -- \\
\(\lambda_N\) & Quantity-oriented PI type & -- \\
\(\mu_\lambda\) & Mass of PIs of type \(\lambda\) & \(\mu_{\lambda_Q}=\mu,\ \mu_{\lambda_N}=1-\mu\) \\
\(\Theta_\lambda\) & Ability set assigned to type-\(\lambda\) PIs & Subset of \([\theta_L,\theta_H]\) \\
\midrule
\multicolumn{3}{l}{\textit{RA characteristics and choices}} \\
\(\theta\) & RA ability & \(\mathrm{TN}(\mu_\theta,\sigma_\theta^2;\theta_L,\theta_H)\) \\
\(e\) & RA effort & \(\mathbb{R}_+\) \\
\(\bar U\) & RA outside option & \(\mathbb{R}\) \\
\(V\) & Continuation value of elite PhD admission & \(\mathbb{R}_+\) \\
\midrule
\multicolumn{3}{l}{\textit{PI choices and technology}} \\
\(K\) & Exogenous AI capability frontier & \(\mathbb{R}_+\) \\
\(n_\lambda\) & Filled normalized team intensity / mass of filled RA positions & \(\mathbb{R}_+\) \\
\(\alpha_\lambda\) & Share of AI allocated to routine tasks & \([0,1]\) \\
\(g_\lambda\) & Mentoring intensity & \(\mathbb{R}_+\) \\
\(K^T_{R,\lambda}\) & Routine-task AI allocation & \(\alpha_\lambda K\) \\
\(K^T_{N,\lambda}\) & Novel-task AI allocation & \((1-\alpha_\lambda)K\) \\
\midrule
\multicolumn{3}{l}{\textit{Human-input terms and task technology}} \\
\(h_R\) & Human input into routine tasks & \(\theta+\eta_R e\) \\
\(h_N\) & Mentored human input into novel tasks & \(\theta+\eta_N e+\psi g_\lambda\) \\
\(A_R(\cdot)\) & Inseparable AI-generated component of routine artifact & Mean zero \\
\(\sigma_A^2(\cdot)\) & Variance of the inseparable AI-generated component & Increasing in \(K_R^T\) \\
\(\sigma_h^2,\sigma_\nu^2\) & Residual human-input variance and artifact-noise variance & \(\mathbb{R}_{++}\) \\
\(\rho_R(\cdot)\) & Derived diagnostic loading of routine evidence & \((0,1]\) \\
\(\kappa_\lambda(\cdot)\) & Novel-task AI augmentation function & \((0,\infty)\) \\
\midrule
\multicolumn{3}{l}{\textit{Outputs, signals, and admissions}} \\
\(y_R,y_N\) & Deterministic routine and novel task outputs & \(\mathbb{R}_+^2\) \\
\(s_R,s_N\) & Market-observed noisy task signals & \(\mathbb{R}^2\) \\
\(\hat{\theta}\) & Linear evaluation of research potential & \(\mathbb{R}\) \\
\(S\) & Admissions-relevant score & \(\mathbb{R}\) \\
\(r_\lambda\) & Reduced-form credibility adjustment for PI environment & \((0,\infty)\) \\
\(S^*\) & Fixed-capacity admissions cutoff & \(\mathbb{R}\) \\
\(P_\lambda(\theta,e;S^*)\) & Admission probability & \([0,1]\) \\
\(Q\) & Mass of elite PhD slots & \(\mathbb{R}_+\) \\
\midrule
\multicolumn{3}{l}{\textit{Research production}} \\
\(Y_\lambda\) & CES research output & \(\mathbb{R}_+\) \\
\(\bar Y_\lambda\) & Expected per-RA CES research output in segment \(\lambda\) & \(\mathbb{R}_+\) \\
\(\varsigma_\lambda\) & Elasticity of substitution between routine and novel outputs & \(\mathbb{R}_+\) \\
\(\bar p_{B,\lambda}\) & Average probability of breakthrough success & \([0,1]\) \\
\midrule
\multicolumn{3}{l}{\textit{Score-load coefficients}} \\
\(A_\lambda^B\) & Deterministic component of the admissions score & \(\mathbb{R}\) \\
\(B_\lambda^B\) & Effort loading in the admissions score & \(\mathbb{R}_+\) \\
\(\Sigma_\lambda^B\) & Residual standard deviation of the admissions score & \(\mathbb{R}_+\) \\
\bottomrule
\end{tabular}
\end{table}

Auxiliary coefficients such as \(A_\lambda^B\), \(B_\lambda^B\), and \(\Sigma_\lambda^B\), as well as the fuller parameterization used in the technical derivations, are reported in the online appendix.

\subsection{Timing and Sequence of the Game}
\label{sec:timing}

Having defined the agents, production technology, signals, and payoffs, we now state the timing of the game. The sequence links the objects above into a single mechanism: laboratory organization affects task output, task output generates noisy evidence, evidence determines admissions scores, and fixed capacity determines placement.

At the beginning of the period, each PI observes the common AI capability frontier \(K\) and chooses laboratory organization,
\[
x_\lambda\equiv(n_\lambda,\alpha_\lambda,g_\lambda).
\]
The PI's AI-related decision is not how much AI exists, but how the available frontier is allocated across routine and novel research tasks.

After the PI has chosen the laboratory environment, a prospective RA in a type-\(\lambda\) segment observes that environment, decides whether to participate, and, if participating, chooses effort \(e\geq0\) to maximize \(U_{RA,\lambda}(\theta,e;S^*)\). Effort affects both research production and evaluation. It enters task output through the routine and novel production functions, and it also affects the noisy task-level signals observed by the admissions market. In equilibrium, the chosen team-intensity policy is interpreted as filled intensity after participation and sorting, not as the number of initially posted slots.

At the end of the period, candidates enter the admissions market with score \(S=r_\lambda\hat{\theta}\). The admissions market sets a cutoff \(S^*\) to allocate the fixed mass \(Q\) of elite PhD slots. Because capacity is fixed in the short run, admissions are determined by relative standing as well as absolute performance.

We characterize equilibrium by backward induction together with the market-clearing condition for \(S^*\). In the final stage, the admissions market clears through the fixed-capacity cutoff. In the second stage, RAs choose effort and decide whether to participate, anticipating how effort affects their admissions score and how the cutoff affects admission chances. In the first stage, PIs choose \(x_\lambda\), anticipating the induced RA effort response, participation decisions, and admissions environment. Individual PIs are atomistic in the continuum economy, so each PI takes \(S^*\) as given even though \(S^*\) is determined in aggregate equilibrium.\footnote{This is the standard price-taking logic in a continuum economy. A single PI has zero mass and does not internalize the effect of their own laboratory choices on the aggregate admissions cutoff \citep{cole1998class,hopkins2012job}.}

\section{Equilibrium Analysis and Propositions}
\label{sec:equilibrium}

For each PI type \(\lambda\in\{\lambda_Q,\lambda_N\}\), an equilibrium consists of an RA effort schedule \(e_\lambda^*(\theta)\), a sorting set \(\Theta_\lambda\), a PI organizational choice
\[
x_\lambda^*\equiv(n_\lambda^*,\alpha_\lambda^*,g_\lambda^*),
\]
and an admissions cutoff \(S^*\). Here \(n_\lambda^*\) denotes filled normalized team intensity after participation and sorting, not posted capacity. The set \(\Theta_\lambda\) is the primitive sorting object: it contains the RA ability types assigned to the type-\(\lambda\) segment. Under the maintained monotonicity and single-crossing conditions, \(\Theta_\lambda\) can be represented by a lower participation cutoff \(\underline{\theta}_\lambda\).\footnote{Without monotonicity or single crossing, \(\Theta_\lambda\) need not be an interval and the cutoff representation may fail. The equilibrium object is therefore the sorting set \(\Theta_\lambda\); \(\underline{\theta}_\lambda\) is a convenient representation in the maintained region.}

The equilibrium has three requirements. First, RAs choose effort and participation optimally, taking the laboratory environment and admissions cutoff as given, as in signaling settings where agents respond to expected evaluation outcomes \citep{spence1973job}. Second, each PI chooses \(x_\lambda^*\) to maximize the type-specific payoff defined in Section~\ref{sec:payoffs}, taking the induced RA behavior and admissions cutoff as given.
Third, the admissions cutoff clears the fixed-capacity PhD market, following the tournament logic that scarce positions are allocated by relative standing \citep{lazear1981rank, hopkins2012job}. Together, these requirements combine individual optimality, organizational optimality, sorting, and market clearing under noisy evaluation \citep{daley2014market}.

The final equilibrium object is the admissions cutoff, which aggregates the outcomes generated by PI choices and RA effort into a capacity-constrained admissions market. Let \(Q\) denote the exogenously fixed measure of available PhD slots. For each PI type \(\lambda\), let \(F_\lambda\) denote the conditional equilibrium distribution of \(\theta\) among RAs who fill positions in the type-\(\lambda\) segment, normalized on \(\Theta_\lambda\). Since \(n_\lambda^*\) is filled normalized team intensity, \(\mu_\lambda n_\lambda^*\) is the aggregate mass of participating candidates from that segment reaching the admissions market. Given the induced PI choices, RA effort schedules, and sorting sets, the cutoff \(S^*\) clears the admissions market:
\begin{equation}
Q
=
\sum_{\lambda\in\{\lambda_Q,\lambda_N\}}
\mu_\lambda n_\lambda^*
\int
P_\lambda\bigl(\theta,e_\lambda^*(\theta);S^*\bigr)\,dF_\lambda(\theta),
\label{eq:eqm_cutoff}
\end{equation}
where \(P_\lambda(\theta,e;S^*)\) is the admission probability defined in Section~\ref{sec:payoffs}. No additional participation-rate multiplier appears because participation and sorting are already embedded in the filled team-intensity convention and in the conditional distribution \(F_\lambda\). If \(n_\lambda\) were instead modeled as posted capacity, the leading mass would need to be replaced by \(M_\lambda^*=\mu_\lambda n_\lambda^*\pi_\lambda^*\), with \(\pi_\lambda^*=\Pr(\theta\in\Theta_\lambda)\). The right-hand side is therefore the total expected mass of admitted RAs across all PI types. For fixed induced PI choices, effort schedules, and sorting sets, this mass is decreasing in \(S^*\). With equilibrium policy responses, the appendix states sufficient local conditions under which the induced aggregate admissions function is continuous, crosses \(Q\), and yields a locally well-defined cutoff.

The equilibrium logic has three main implications. First, routine-task AI can increase routine output while reducing the diagnostic precision of routine evidence about human contribution; in the maintained parameter region, this also lowers the linear weight placed on routine evidence in the evaluation rule. Second, heterogeneous PI objectives and task complementarities can generate different laboratory organizations. Third, fixed admissions capacity can turn broad improvements in candidate records into a cutoff effect: when scores rise widely and the candidate-mass response does not offset the shift, the threshold for admission rises. The remainder of this section formalizes these implications as propositions.

\subsection{Proposition 1: Routine-Output Expansion and Reduced Informativeness}
\label{sec:prop1}

The first proposition formalizes the paper's central informational tension. AI can make routine work easier to produce, but the same routine output may become less informative about the RA's own ability and effort. This distinction is important because pre-doctoral work is both productive labor and evidence for later evaluation \citep{autor2003skill, noy2023experimental, brynjolfsson2025generative, doshi2024generative, dellacqua2026jagged}.
\begin{proposition}[Routine-output expansion and reduced informativeness]
\label{prop:dual_ai_signal}
Fix total AI capability \(K\), and let \(x\equiv K_R^T=\alpha K\) denote routine-task AI intensity. Under the routine production function in (1) and the routine signal in (2), suppose \(a_R'(x)>0\), \(m_R'(x)\geq 0\), and \(h_R>0\). If the diagnostic loading \(\rho_R(x)\) is generated by the artifact-level attribution problem described above, then \(\rho_R'(x)<0\). Holding the local human-input term \(h_R\) fixed:

\begin{enumerate}
    \item[(i)] routine-task AI weakly increases routine output:
    \[
    \frac{dy_R(x)}{dx}
    =
    a_R'(x)+m_R'(x)h_R
    \geq 0;
    \]

    \item[(ii)] routine-task AI reduces the diagnostic precision of routine evidence about the RA's human contribution:
    \[
    \frac{d}{dx}
    \left(
    \frac{\rho_R(x)^2}{\sigma_R^2}
    \right)
    =
    \frac{2\rho_R(x)\rho_R'(x)}{\sigma_R^2}
    <0;
    \]

    \item[(iii)] as routine evidence ceases to load on human contribution, \(\rho_R(x)\to0\), the linear prediction coefficient on routine evidence also converges to zero:
    \[
    \omega_R^B(\alpha,K,\lambda)
    =
    \frac{\rho_R(\alpha K)/\sigma_R^2}
    {
    \frac{1}{\sigma_\theta^2}
    +
    \frac{\rho_R(\alpha K)^2}{\sigma_R^2}
    +
    \frac{\kappa_\lambda((1-\alpha)K)^2}{\sigma_N^2}
    }
    \to0.
    \]
\end{enumerate}

Thus, routine-task AI can raise routine output while reducing the usefulness of routine evidence for evaluating the RA's own human contribution and, in the residualized benchmark, the RA's underlying ability.
\end{proposition}

Proposition~\ref{prop:dual_ai_signal} is a partial comparative static: it isolates the routine-AI allocation margin before equilibrium effort, sorting, and the admissions cutoff adjust. The result shows that routine output and routine evidence can move in opposite directions. Routine artifacts may become more abundant and more polished, while carrying less diagnostic information about the RA's own contribution. The loss of diagnostic content is not assumed directly; it follows from the projection problem created by an inseparable AI-generated artifact component.

This does not imply that evaluation simply shifts to a better signal. Novel contribution, such as judgment, originality, and interpretation, may be more meaningful but also harder to assess ex ante than standardized execution. Evidence from research funding evaluation shows that structurally diverse teams can be penalized in ex ante review even when they perform better ex post \citep{banalestanol2019evaluation}.\footnote{Grant review and pre-doctoral admissions are different institutions. The analogy is limited to the evaluation problem: complex or nonstandard contributions may be valuable ex post but difficult to assess ex ante.} This suggests that, if AI weakens routine evidence, evaluators need better ways to assess complex research contribution. The mechanism is also consistent with evidence that generative AI can raise productivity while compressing observed performance differences or reducing output diversity in structured knowledge tasks \citep{noy2023experimental,brynjolfsson2025generative,doshi2024generative,dellacqua2026jagged,tambe2025reskilling}.

\paragraph{Implications.}

Three immediate implications follow from Proposition~\ref{prop:dual_ai_signal}. First, visible output can improve while the overall precision of evaluation falls if the loss of routine precision is not offset by more informative novel evidence. In the residualized linear-Gaussian benchmark, total precision is
\[
\frac{1}{\sigma_\theta^2}
+
\frac{\rho_R(K^T_{R,\lambda})^2}{\sigma_R^2}
+
\frac{\kappa_\lambda(K^T_{N,\lambda})^2}{\sigma_N^2}.
\]
When routine-task AI lowers \(\rho_R(K^T_{R,\lambda})\), the routine component of precision falls. Second, independent process evidence becomes more valuable when baseline precision is lower. Third, the admissions return to effort shifts toward the signal channels that remain diagnostic. These implications reflect the same wedge: the AI allocation that increases research throughput need not be the allocation that best reveals candidate ability. The online appendix reports the corresponding precision and effort-loading algebra.

\begin{figure}[t]
\centering
\includegraphics[width=\linewidth]{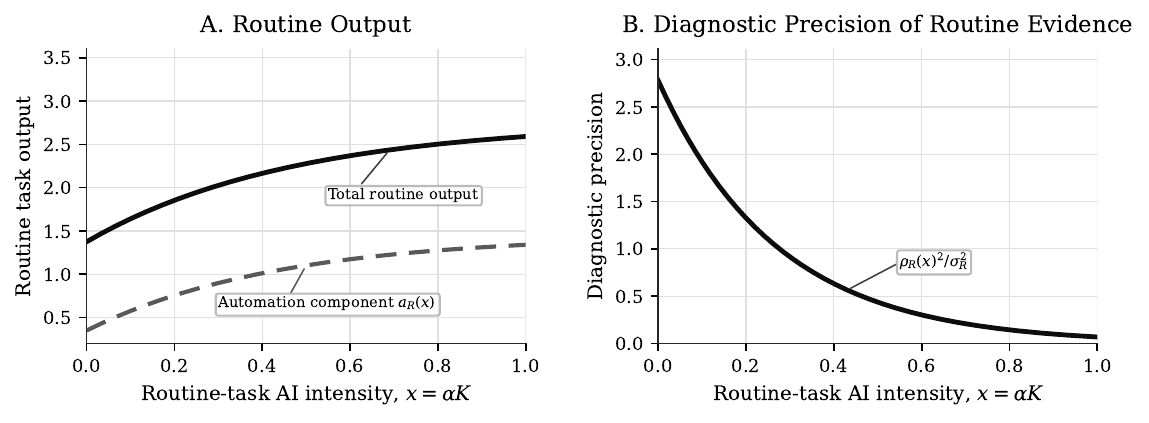}
\caption{Routine-output expansion and diagnostic compression. Panel A shows that, holding human input fixed, higher routine-task AI intensity raises the automation component and total routine output. Panel B shows that the diagnostic precision of routine evidence about human contribution, \(\rho_R(x)^2/\sigma_R^2\), declines as routine-task AI adds inseparable AI-generated variation to routine artifacts and thereby lowers the derived loading on human contribution. This figure is illustrative and uses the parameterization reported in the online appendix. Panel B uses a smooth numerical path consistent with the derived projection loading from the artifact-level signal-extraction microfoundation in the online appendix's Assumption A4, not a separate imposed sign restriction on \(\rho_R'(x)\).}
\label{fig:signal_compression_main}
\end{figure}

\begin{figure}[h!]
\centering
\includegraphics[width=\linewidth]{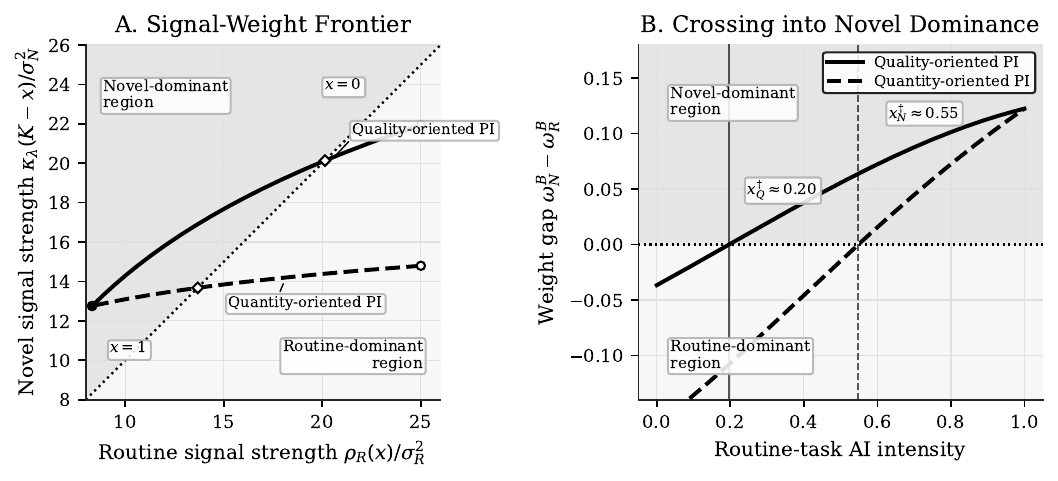}
\caption{
Illustrative signal reweighting as routine evidence loses diagnostic content. 
Panel A plots the signal-weight frontier in the space of routine and novel signal strength. The horizontal axis is the routine signal strength, \(\rho_R(x)^2/\sigma_R^2\), and the vertical axis is the novel signal strength, \(\kappa_\lambda(K-x)^2/\sigma_N^2\). The dotted diagonal marks the equal-weight boundary, \(\omega_R^B=\omega_N^B\). Points below the boundary correspond to routine-dominant evaluation, where routine evidence receives more weight; points above the boundary correspond to novel-dominant evaluation, where novel-task evidence receives more weight. The solid and dashed curves illustrate how the relative signal strengths move as routine-task AI intensity \(x\) changes for quality-oriented and quantity-oriented PI environments. 
Panel B plots the corresponding weight gap, \(\omega_N^B-\omega_R^B\), against routine-task AI intensity \(x\). The horizontal dotted line at zero marks the equal-weight benchmark, \(\omega_N^B=\omega_R^B\). Values below this line indicate routine-dominant evaluation, where routine evidence receives more weight than novel-task evidence. Values above this line indicate novel-dominant evaluation, where novel-task evidence receives more weight than routine evidence. The solid curve traces the weight-gap path for the quality-oriented PI environment, and the dashed curve traces the path for the quantity-oriented PI environment. As \(x\) increases, routine evidence loses diagnostic content, so both curves move upward toward greater relative weight on novel-task evidence. The vertical solid line marks the illustrative crossing point \(x_Q^*=0.20\) for the quality-oriented environment, where its evaluation rule switches from routine-dominant to novel-dominant weighting. The vertical dashed line marks the corresponding crossing point \(x_N^*=0.55\) for the quantity-oriented environment. The earlier crossing of the solid curve indicates that the quality-oriented environment shifts toward novel-task evidence at a lower level of routine-task AI intensity. The shaded regions label the two regimes: the lower region is routine-dominant, and the upper region is novel-dominant. The figure uses a stylized parameterization and is intended only to visualize the signal-reweighting implication of Proposition~\ref{prop:dual_ai_signal}; it should not be read as an estimated weighting surface.
}
\label{fig:screening_frontier_main}
\end{figure}

Fig.~\ref{fig:screening_frontier_main} provides an illustrative visualization of the signal-reweighting implication of Proposition~\ref{prop:dual_ai_signal}. As routine-task AI reduces the diagnostic content of routine evidence, the linear evaluation rule places less weight on routine signals and relatively more weight on novel-task evidence. The figure is included only to illustrate how evaluation can move from routine-dominant to novel-dominant weighting as routine evidence loses diagnostic content; it is not used in the proof and should not be read as an estimated weighting surface.

\subsection{Proposition 2: Local Characterization of Organizational Segmentation}
\label{sec:prop2}

The second result explains why laboratories need not respond to AI in the same way. PI objectives differ, and routine and novel tasks may be more substitutable in some research environments than in others. The result should be read as a local characterization, not as a global monotone comparative static. Its role is to show how primitive differences in objectives and task technology can generate segmented laboratory strategies \citep{manso2011incentives, garicano2000hierarchies, kremer1993oring}.

\begin{proposition}[Local segmentation under CES task production]
\label{prop:segmentation_revised}
Let \(\Pi_N(n,\alpha,g)\) and \(\Pi_Q(n,\alpha,g)\) denote the type-specific PI payoff functions defined in Section~\ref{sec:payoffs}, where \(\lambda_N\) is the quantity-oriented type and \(\lambda_Q\) is the quality-oriented type. Fix total AI capability \(K\), and consider a common interior comparison point
\[
\bar x=(\bar n,\bar\alpha,\bar g),
\qquad
\bar n>0,\quad \bar\alpha\in(0,1),\quad \bar g>0.
\]
All derivatives below are evaluated at \(\bar x\), under the local-envelope convention that holds fixed the induced participant distribution and the admissions cutoff response.

Suppose that, in a neighborhood of \(\bar x\), each type-specific payoff function is twice continuously differentiable, locally concave in \((n,\alpha,g)\), and has an isolated interior optimum. Suppose also that local cross-choice interactions are weak enough that the sign of each own marginal payoff at \(\bar x\) determines the direction of the corresponding local optimum.\footnote{
A sufficient set of regularity conditions is that the Hessian of each type-specific payoff function with respect to \((n,\alpha,g)\) is negative definite and locally diagonally dominant around \(\bar x\), so that own-curvature terms dominate cross-choice interactions. The appendix provides the corresponding local implicit-function argument and the primitive threshold conditions under which the displayed marginal payoff signs hold.
}
Suppose further that the two PI types satisfy the following local payoff-gradient conditions:
\[
\frac{\partial \Pi_N}{\partial \alpha}(\bar x)>0>
\frac{\partial \Pi_Q}{\partial \alpha}(\bar x),
\tag{A}
\]
\[
\frac{\partial \Pi_Q}{\partial g}(\bar x)>0>
\frac{\partial \Pi_N}{\partial g}(\bar x),
\tag{G}
\]
and
\[
\frac{\partial \Pi_N}{\partial n}(\bar x)\geq 0\geq
\frac{\partial \Pi_Q}{\partial n}(\bar x).
\tag{N}
\]
Then there exists a local equilibrium neighborhood around \(\bar x\) in which
\[
\alpha_N^*>\alpha_Q^*,
\qquad
g_Q^*>g_N^*,
\qquad
n_N^*\geq n_Q^*.
\]
If the scale condition in \((N)\) holds strictly, then \(n_N^*>n_Q^*\).
\end{proposition}

The economic interpretation is straightforward. Quantity-oriented PIs place greater value on scalable CES research output. When routine and novel tasks are sufficiently substitutable, routine-task automation has a stronger local payoff value for these PIs because it supports scalable research throughput. Quality-oriented PIs place greater value on upper-tail research success. When routine and novel tasks are sufficiently complementary, shifting AI capacity toward routine tasks is less attractive because routine execution cannot easily substitute for novel-task augmentation, research design, interpretation, and mentoring.

The team-size comparison requires an additional local scale condition. The value of upper-tail success must be high enough to make mentoring and novel-task preservation attractive for quality-oriented PIs, but not so high, relative to coordination costs and portfolio diminishing returns, that quality-oriented PIs also expand normalized team intensity at the comparison point. Under these primitive conditions, the local payoff-gradient signs in Proposition~\ref{prop:segmentation_revised} hold. The online appendix derives the corresponding threshold conditions in terms of the CES derivatives, breakthrough-probability derivatives, and organizational cost primitives. These are sufficient local conditions, not universal predictions for arbitrary AI shocks.\footnote{A stronger global monotone-comparative-statics result could be obtained under additional increasing-differences and supermodularity assumptions. The paper uses the weaker local characterization because it is sufficient for the baseline mechanism.}

This maps naturally onto academic incentive environments in which some rewards concentrate around high-impact research outcomes while others reward sustained measurable output \citep{heckman2020publishing, hamermesh2013six, edwards2017academic, hicks2012performance}. A continuous distribution of PI objectives would replace the two-type partition with a continuum of organizational choices; the result should then be read as identifying local sorting directions along that continuum rather than as requiring only two discrete laboratory types.

\begin{figure}[t]
\centering
\includegraphics[width=\linewidth]{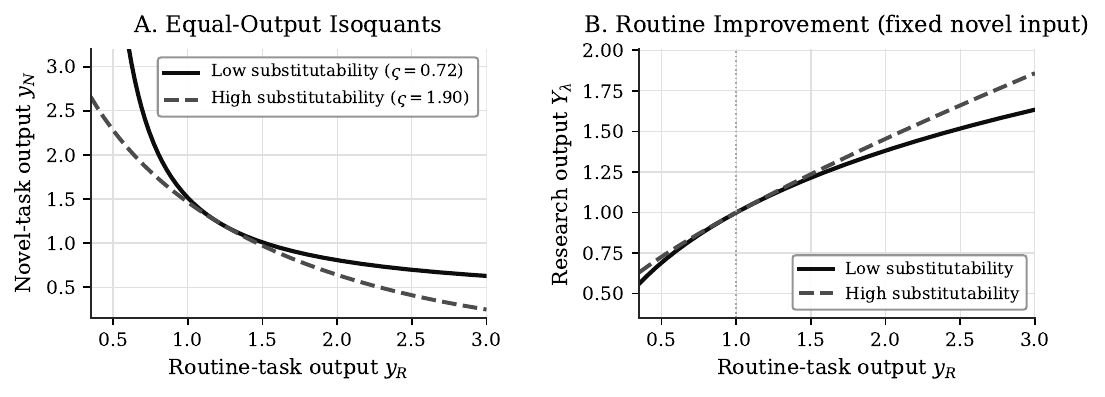}
\caption{CES task environments. Panel A plots equal-output isoquants for low- and high-substitutability environments. Lower substitutability implies stronger complementarity between routine and novel tasks. Panel B holds novel-task input fixed and shows that routine-task improvements translate more strongly into research output when routine and novel tasks are more substitutable. This figure is illustrative and uses the parameterization reported in the online appendix; it visualizes CES task environments rather than estimated production elasticities.}
\label{fig:ces_regimes_main}
\end{figure}

The equilibrium implication is organizational segmentation. One segment uses AI to scale structured research production; the other relies more on mentoring and novel-task support. The model does not claim that this segmentation is absolute in practice. It shows that such segmentation can arise from heterogeneous PI objectives interacting with task complementarity and AI allocation. Fig.~\ref{fig:ces_regimes_main} visualizes the production-side mechanism behind this result \citep{fudenberg2019training, kostadinov2022learning, garicano2000hierarchies}.

\begin{table}[h!]
\centering
\caption{Characterization of Equilibrium Strategies in the Maintained Local Region}
\label{tab:equilibrium_strategies_revised}
\begin{threeparttable}
\begin{tabular}{>{\raggedright\arraybackslash}p{3.2cm}>{\raggedright\arraybackslash}p{5.0cm}>{\raggedright\arraybackslash}p{5.0cm}}
\toprule
\textbf{Strategy Component} & \textbf{Quality-Oriented PI (\(\lambda_Q\))} & \textbf{Quantity-Oriented PI (\(\lambda_N\))} \\
\midrule
Primary Objective &
Maximize the probability of at least one high-value novel outcome, with stronger task complementarity &
Maximize scalable CES research output, with higher effective task substitutability \\
\midrule
AI Allocation \((\alpha^*)\) &
Lower routine-task share; stronger tilt toward novel-task augmentation &
Higher routine-task share; stronger tilt toward automation / scaling \\
\midrule
Mentoring \((g^*)\) &
Higher mentoring intensity; guidance complements novel-task performance &
Lower mentoring intensity on average; guidance is relatively more costly at scale \\
\midrule
Normalized Team Intensity \((n^*)\) &
Lower normalized team intensity; value concentrated in intensively developed RAs &
Weakly higher normalized team intensity; pipeline production becomes more attractive \\
\midrule
Implied RA Skill Focus &
Idea generation, judgment, and mentored problem-solving &
Workflow management, structured execution, and scalable production \\
\bottomrule
\end{tabular}
\begin{tablenotes}[flushleft]
\item Notes: The entries summarize the maintained local parameter region of Proposition~\ref{prop:segmentation_revised}. They are not universal predictions for all laboratories, estimated frequencies, or structural parameter estimates.
\end{tablenotes}
\end{threeparttable}
\end{table}

\subsection{Proposition 3: Fixed-Capacity Congestion}
\label{sec:prop3}

The third proposition states the fixed-capacity result. The key idea is simple: if many laboratories generate stronger candidate records at the same time, and elite admissions capacity remains fixed, some of the gain can be absorbed by a higher admissions cutoff rather than by a proportional expansion of opportunity \citep{lazear1981rank, hopkins2012job, card2013nine}.

\begin{proposition}[Location-shift congestion under fixed capacity]
\label{prop:arms_race_revised}
Let \(M(\bar K)\equiv\sum_{\lambda}\mu_\lambda n_\lambda^*(\bar K)\) be the equilibrium candidate mass reaching the admissions market, and suppose \(M(\bar K)>Q\). Suppose aggregate AI capability shifts admissions scores by
\[
S_i(\bar K)=\Delta(\bar K)+u_i,
\qquad
\Delta'(\bar K)>0,
\]
where \(u_i\sim H_0\) and \(H_0\) has density \(h_0>0\) at the relevant upper quantile. The fixed-capacity cutoff is
\[
S^*(\bar K)
=
\Delta(\bar K)
+
H_0^{-1}\!\left(1-\frac{Q}{M(\bar K)}\right).
\]
For fixed \(Q\), defining
\[
q(\bar K)=H_0^{-1}\!\left(1-\frac{Q}{M(\bar K)}\right),
\]
we have
\[
\frac{dS^*(\bar K)}{d\bar K}
=
\Delta'(\bar K)
+
\frac{Q M'(\bar K)}
{M(\bar K)^2 h_0(q(\bar K))}.
\]
Thus, the cutoff rises whenever \(M'(\bar K)\geq0\), and entry amplifies the cutoff effect. If capacity also changes with \(\bar K\), then
\[
\frac{dS^*(\bar K)}{d\bar K}
=
\Delta'(\bar K)
+
\frac{
Q(\bar K)M'(\bar K)/M(\bar K)^2
-
Q'(\bar K)/M(\bar K)
}
{h_0(q(\bar K))},
\]
so capacity expansion dampens the cutoff effect. For any candidate whose own score components do not rise enough to offset the higher standardized cutoff, admission probability falls. Hence, under fixed or slowly adjusting capacity, part of the private placement gain from AI is competed away through a congestion channel.
\end{proposition}

Proposition~\ref{prop:arms_race_revised} is still a conditional statement, but the condition is now primitive: AI shifts the market score distribution by a common component \(\Delta(\bar K)\). It does not say that every AI shock must shift the entire score distribution to the right, nor that AI necessarily lowers admission probabilities. It says that when AI produces a broad location shift in visible records, a fixed-capacity market converts that shift one-for-one into a higher cutoff; if more candidates enter the market, the relevant upper quantile moves further right. If capacity expands, or if candidate mass falls, the cutoff effect can be weakened. This is the same logic that makes rank-based competition different from evaluation against a fixed absolute standard \citep{lazear1981rank, hopkins2023is}. It is also related to evidence and models of escalating standards in academic and publication markets \citep{card2013nine, ellison2002slowdown}.

The welfare implication is also conditional. AI may raise research productivity, candidate learning, and the quality of visible records. Proposition~\ref{prop:arms_race_revised} isolates one opposing force: in a scarce admissions market, part of the private signaling gain can be competed away. The result is therefore a congestion mechanism, not a claim that AI is socially harmful.

Section~\ref{sec:model_faithful_simulation} below reports a mechanism-preserving simulation that places this cutoff logic in a denser stochastic environment. Following the transparent simulation logic of \citet{kapeller2016emergent}, the exercise is illustrative rather than structural: it uses the model's own task-output equations, signal weights, PI payoff, and fixed-capacity admissions rule while drawing RA talent, PI orientation, project luck, and assessment noise at high Monte Carlo density.

\begin{figure}[t]
\centering
\includegraphics[width=\linewidth]{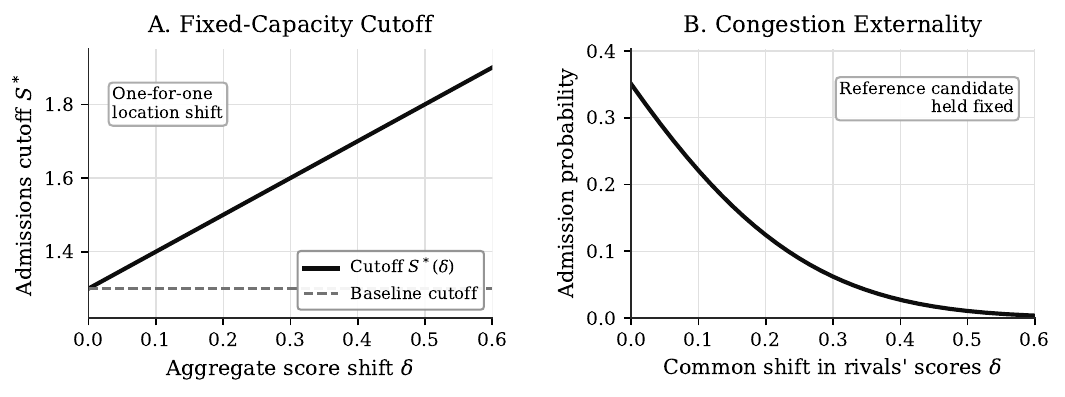}
\caption{Congestion under fixed admissions capacity. Panel A shows a simple location-shift benchmark: when the aggregate score distribution shifts upward and admissions capacity is fixed, the cutoff rises. Panel B holds a reference candidate's own score distribution fixed while rivals' scores shift upward; the rising cutoff lowers the reference candidate's admission probability. This figure is illustrative and uses a stylized location-shift benchmark; it is not a structural estimate of admissions cutoffs.}
\label{fig:congestion_main}
\end{figure}

Fig.~\ref{fig:congestion_main} visualizes the central comparative static of Proposition~\ref{prop:arms_race_revised}. In a fixed-capacity tournament, when many candidates receive a common upward shift in admissions scores, the cutoff rises with the shifted score distribution. This is the standard rank-order logic of tournament allocation under scarce positions \citep{lazear1981rank, hopkins2012job, hopkins2023is}. The figure also shows the associated congestion effect: a reference candidate whose own score distribution is held fixed can face a lower admission probability when rivals' scores rise and the admissions cutoff moves upward.

\begin{table}[t]
\centering
\caption{Dominant Comparative-Static Directions in the Maintained Parameter Region}
\label{tab:comparative_statics_summary}
\begin{threeparttable}
\scriptsize
\setlength{\tabcolsep}{3pt}
\begin{adjustbox}{max width=\linewidth,center}
\begin{tabular}{>{\raggedright\arraybackslash}p{3.0cm}>{\raggedright\arraybackslash}p{2.1cm}>{\raggedright\arraybackslash}p{1.9cm}>{\raggedright\arraybackslash}p{1.9cm}>{\raggedright\arraybackslash}p{2.0cm}>{\raggedright\arraybackslash}p{2.0cm}}
\toprule
\textbf{AI-environment change} & \(\boldsymbol{\alpha^*}\) & \(\boldsymbol{g^*}\) & \(\boldsymbol{n^*}\) & \textbf{Routine evidence} & \(\boldsymbol{S^*}\) \\
\midrule
Stronger routine-task automation & Higher in scalable / quantity-oriented segments & Lower on average where routine scaling substitutes for intensive guidance & Higher in scalable segments & Lower diagnostic precision when \(\rho_R\) falls; lower \(\omega_R^B\) in the maintained region & Higher under a market-wide capability shift; otherwise no direct market-wide effect \\
\midrule
Stronger novel-task augmentation & Lower, especially in quality-oriented segments & Higher & Often smaller, more intensive quality-oriented teams & Lower relative weight as novel signals become more informative & Higher under a market-wide capability shift; otherwise no direct market-wide effect \\
\midrule
Stronger compression in observable routine performance & Ambiguous for production, but less evidentiary value from routine tilt & Higher in evaluation-intensive segments & Ambiguous & Diagnostic precision and routine weight lower & Higher if many candidates bunch in the upper tail \\
\midrule
Higher task substitutability \((\varsigma_\lambda)\) & Higher, especially in pipeline-style environments & Lower & Higher & No direct effect & No direct effect except through induced equilibrium composition \\
\bottomrule
\end{tabular}
\end{adjustbox}
\begin{tablenotes}[flushleft]
\item Notes: The entries report dominant directions from the maintained parameter region rather than universal global comparative statics. The table is a qualitative map of analytical comparative statics, not a calibrated simulation table. Cells marked ``ambiguous'' indicate that the sign depends on PI type, on whether the shock is idiosyncratic or aggregate, or on whether production incentives or evaluation incentives are the relevant margin. The robust diagnostic-compression result in Proposition~\ref{prop:dual_ai_signal} is the decline in the human-contribution diagnostic precision \(\rho_R^2/\sigma_R^2\); the decline in \(\omega_R^B\) is reported for the maintained parameter region. The cutoff effect follows Proposition~\ref{prop:arms_race_revised} and applies to broad market-wide score shifts under fixed capacity.
\end{tablenotes}
\end{threeparttable}
\end{table}

\subsection{Extension 1: Apprenticeship Learning and Human Capital Accumulation}
\label{sec:learning_extension}

The baseline focuses on evaluation and congestion, but pre-doctoral work may also build human capital. The online appendix captures this apprenticeship channel with a recursive extension in which future research capability rises with effort, novel-task AI exposure, and mentoring \citep{becker1962investment, fudenberg2019training, kostadinov2022learning}. The extension distinguishes a learning-dominant regime, where mentoring and difficult task exposure build durable capability, from a signaling-dominant regime, where effort mainly improves current admissions scores and may therefore reflect tournament congestion or signal escalation \citep{hopkins2023is}.

\subsection{Extension 2: Social Value, Knowledge Production, and Welfare}
\label{sec:welfare_extension}

The welfare extension separates private ranking incentives from the broader social value of research. The online appendix decomposes welfare into knowledge output, human-capital gains, effort costs, and mentoring costs. AI can raise social value by increasing research output and building future capability when paired with mentoring \citep{becker1962investment, manso2011incentives}. But in a fixed-capacity tournament, some private gains may be absorbed by higher cutoffs and signal escalation \citep{lazear1981rank}. The CES structure matters because routine-task gains have greater social value when routine and novel tasks are complements rather than substitutes for judgment-intensive work \citep{kremer1993oring}. Therefore, the welfare effect is ambiguous: the model identifies when AI-generated productivity gains become social value and when they are partly converted into competitive signaling.

\section{Mechanism-Preserving Simulation with Ability, Luck, and Bottlenecked Selection}
\label{sec:model_faithful_simulation}

\subsection{Simulation Design and Purpose}

The analytical model isolates the production--evaluation mechanism in closed form. This section adds an agent-level simulation. The simulation is designed as a robustness exercise for the model's three mechanisms. 
It examines whether the model’s qualitative mechanisms continue to hold in a richer setting with heterogeneous RAs and PIs, luck in research outcomes, noisy admissions evaluation, and a fixed number of elite PhD slots.
This simulation follows a tradition that uses simple agent-level models to examine how evaluation rules and institutional bottlenecks can generate aggregate selection patterns that are not obvious from individual-level assumptions \citep{ahrweiler2015modelling, kapeller2016emergent}.

The simulation therefore examines three mechanism-level claims. First, routine-task AI can raise routine output while reducing the diagnostic content of routine evidence. Second, PI organizational choices can segment over a continuous research-orientation space rather than only across two discrete types. Third, fixed capacity can convert noisy evaluation and project luck into false negatives among high-ability or high-realized-merit candidates. The last margin is motivated by work showing that observed success can combine talent with randomness, and that scarce evaluation systems can turn noisy assessments into misallocation \citep{biondo2018talent, kapeller2016emergent}.

Parameter values are chosen to produce an informative baseline environment. Admission is neither nearly universal nor nearly impossible; PI choices vary across the relevant range; and admissions capacity is scarce enough for ranking errors to affect selection outcomes.
The online appendix reports the full parameter table and sensitivity checks over alternative capacity ratios, uncertainty scales, diagnostic-compression strengths, effort rules, assignment rules, and a finer \(\lambda\)-grid. 
We make two simplifying choices to keep the simulation focused. First, as \(K\) changes, we keep the number of candidates and the admissions capacity ratio fixed, so the simulation captures changes in scores and cutoffs rather than changes in market entry. Second, we do not simulate a full RA--PI matching market. Instead, RAs are assigned to lab environments in proportion to the filled RA positions in those environments, while the analytical model provides the formal sorting condition through \(\Theta_\lambda\).
The appendix therefore reports a stronger rank-sorted assignment check that keeps the same filled-position masses but sorts RAs by comparative novel-versus-routine talent across the PI-orientation continuum.

\subsection{Agents, Production, and Evaluation}

Each replication contains \(J\) PIs and \(I\) RAs. PI \(j\) draws a continuous research-orientation index
\[
\lambda_j\sim U[0,1],
\]
where \(\lambda_j=0\) corresponds to the quantity-oriented endpoint and \(\lambda_j=1\) corresponds to the quality-oriented endpoint. Each type-specific primitive \(v\in\{\phi,\varsigma,r,\Omega,\gamma\}\) is interpolated between the analytical endpoints:
\[
v(\lambda_j)=(1-\lambda_j)v_N+\lambda_j v_Q.
\]
Thus, the two-type analytical model should be read as an endpoint representation of a continuum. To make quality-oriented work differ in more than subjective payoff weights, the simulation also lets project uncertainty and reception uncertainty rise with the novelty-oriented endpoint:
\[
\sigma_\zeta(\lambda_j)=\sigma_{\zeta,0}(1+a_\zeta\lambda_j),
\qquad
\sigma_\xi(\lambda_j)=\sigma_{\xi,0}(1+a_\xi\lambda_j).
\]
The interpretation is that more exploratory research can have higher upper-tail potential, but also higher project risk and noisier ex ante assessment.
This maps the simulation to work on innovation incentives, unexpected research outcomes, and AI-assisted scientific search, where novelty is valuable partly because its payoffs are uncertain and upper-tailed \citep{manso2011incentives, aslan2024unexpectedness, agrawal2024artificial}.
This is an exploratory-research specification, not a universal claim that novelty always increases evaluation noise. Settings with better documentation, audit trails, or process evidence could have flatter \(\sigma_\xi(\lambda)\) schedules.

RA talent is distributed and multidimensional. Each RA draws routine-execution talent and novel-research talent,
\[
\begin{pmatrix}
\theta_i^R\\
\theta_i^N
\end{pmatrix}
\sim
TN_2(\mu_\theta,\Sigma_\theta;[\theta_L,\theta_H]^2),
\]
with positive but imperfect correlation. The simulation summarizes latent ability as
\[
\theta_i^{L}=\frac{\theta_i^R+\theta_i^N}{2}.
\]
Human input is
\[
h^R_{ij}=\theta_i^R+\eta_R e_{ij},
\qquad
h^N_{ij}=\theta_i^N+\eta_N e_{ij}+\psi g_j.
\]
For computational transparency, effort is represented by a monotone effort-loading approximation based on the analytical score loading:
\[
e_{ij}^*
=
\max\{0,\,
\beta_{RA}\chi(\theta_i^L,\alpha_jK)\,B_j^B\},
\]
where
\[
B_j^B
=
r(\lambda_j)\left[
\omega_R^B\rho_R(K^T_{R,j})\eta_R
+
\omega_N^B\kappa_{\lambda_j}(K^T_{N,j})\eta_N
\right].
\]
This keeps the simulation tied to the model's effort channel while allowing realized effort to vary with ability and the PI environment. 
This effort rule is a computational approximation rather than a full solution to the RA's cutoff-dependent effort problem in \eqref{eq:e_star}. The analytical model and the appendix provide the formal effort-choice problem; the simulation uses this simpler rule to preserve the main effort margin without adding another equilibrium fixed-point calculation.
As a check, the online appendix replaces this rule with a cutoff-sensitive best-response approximation: given a simulated cutoff, each RA maximizes expected admission value net of quadratic effort cost over a one-dimensional effort grid, and the cutoff is updated from the induced score distribution. The qualitative selection-error results are unchanged.

Given the PI's choices \(x_j=(n_j,\alpha_j,g_j)\), task allocations are \(K^T_{R,j}=\alpha_jK\) and \(K^T_{N,j}=(1-\alpha_j)K\). Task outputs follow the main model,
\[
y^R_{ij}=a_R(K^T_{R,j})+m_R(K^T_{R,j})h^R_{ij},
\qquad
y^N_{ij}=\kappa_{\lambda_j}(K^T_{N,j})h^N_{ij},
\]
and are aggregated into \(Y_{ij}\) using the CES research technology. PIs choose \(x_j\) by grid search over the analytical PI payoff on a scalar latent-ability grid. This policy-precomputation step keeps the organizational choice close to the closed-form model, but it is not a full optimization over multidimensional RA matching. RAs are then assigned to PI environments with probability proportional to filled team intensity \(n_j\), the simulation analogue of the filled-position convention in the equilibrium definition. This assignment rule deliberately avoids imposing ability-based sorting mechanically in the baseline simulation. The appendix reports a rank-sorted assignment robustness check that assigns higher comparative novel talent, \(\theta_i^N-\theta_i^R\), to higher-\(\lambda\) PI environments while preserving the same filled team-intensity masses.

The key measurement extension separates three objects that coincide only in the most stylized benchmark. Latent ability \(\theta_i^L\) captures underlying research potential. Realized true merit is
\[
M_{ij}^{\mathrm{true}}
=
Y_{ij}+\zeta_{ij},
\qquad
\zeta_{ij}\sim N(0,\sigma_\zeta(\lambda_j)^2),
\]
where \(\zeta_{ij}\) captures project-level luck and discovery uncertainty. Admissions committees do not observe \(M_{ij}^{\mathrm{true}}\). They observe noisy routine and novel signals, form the same linear evaluation \(\hat\theta_{ij}\) as in the model, and receive a noisy perceived score
\[
S_{ij}^{\mathrm{obs}}
=
r(\lambda_j)\hat\theta_{ij}+\xi_{ij},
\qquad
\xi_{ij}\sim N(0,\sigma_\xi(\lambda_j)^2).
\]
This separation follows stochastic production and noisy-signal models in which latent quality, realized output, and evaluated evidence need not coincide \citep{just1978stochastic, daley2014market, heinsalu2018dynamic, bao2021signal}. The extra term \(\xi_{ij}\) is a simulation-only reception shock layered on top of task-level signal noise; it is included to study assessment uncertainty, not to change the analytical propositions. Admissions are allocated to the top \(Q\) candidates by \(S_{ij}^{\mathrm{obs}}\), while ex post selection quality is evaluated using \(M_{ij}^{\mathrm{true}}\). The top-\(Q\) rule follows rank-order tournament logic under scarce positions \citep{lazear1981rank, hopkins2012job, hopkins2023is}. 
This distinction brings the bottleneck logic of \citet{kapeller2016emergent} into the admissions setting: when scarce slots are allocated using noisy evaluations, some high-ability candidates and high-realized-merit work may be overlooked.

\subsection{Decomposition Benchmarks}

The simulation reports both ability-based and merit-based selection errors:
\[
FN^{ability}
=
\Pr(A_i=0\mid \theta_i^{L}\ge q_{0.8}^{\theta}),
\qquad
FN^{merit}
=
\Pr(A_i=0\mid M_i^{\mathrm{true}}\ge q_{0.8}^{M}),
\]
where \(A_i=1\) denotes admission, \(q_{0.8}^{\theta}\) is the 80th percentile of latent ability, and \(q_{0.8}^{M}\) is the 80th percentile of realized true merit. Because the baseline capacity ratio is \(Q/M=0.17\), which is below the top-quintile threshold, false negatives among top-quintile candidates are mechanically positive even under a highly informative ranking rule. The informative quantity is therefore the change relative to the idealized benchmark, not the level alone.

The enhanced run uses \(K=0.00,0.10,\ldots,2.00\), 80 Monte Carlo replications at each scenario--\(K\) point, 4,000 RAs per replication, 200 PIs per replication, and 31 grid points for continuous-\(\lambda\) policy precomputation. The no-compression benchmark uses the same data-generating distributions for agent populations, PI orientations, project shocks, and the capacity rule while setting \(\rho_R(K^T_R)=1\).

\begin{table}[t]
\centering
\caption{Ability, luck, and evaluation-noise decomposition at \(K=2\)}
\label{tab:abm_decomposition}
\begin{adjustbox}{width=\linewidth}
\begin{tabular}{lccccccc}
\toprule
Scenario & Compression & Luck & Noise & \(\mathrm{corr}(S^{obs},\theta^L)\) & \(FN^{ability}\) & \(FN^{merit}\) & Main channel\\
\midrule
Idealized & No & No & No & 0.725 & 0.501 & 0.757 & \makecell[l]{Fixed-capacity\\benchmark}\\
Luck only & No & Yes & No & 0.724 & 0.503 & 0.781 & \makecell[l]{Ability--merit\\realization risk}\\
Evaluation noise only & No & No & Yes & 0.626 & 0.558 & 0.764 & \makecell[l]{Score--ability\\reception risk}\\
Full noisy AI & Yes & Yes & Yes & 0.510 & 0.617 & 0.838 & \makecell[l]{All channels with\\diagnostic compression}\\
\bottomrule
\end{tabular}
\end{adjustbox}
\begin{minipage}{0.96\linewidth}
\footnotesize
Notes: All rows use \(K=2\) and fixed capacity \(Q/M=0.17\). The idealized row removes the simulation's added diagnostic-compression, project-luck, and reception-noise layers while retaining the model's task-evidence structure, heterogeneous PI environments, and fixed-capacity tournament. It is therefore a model-only benchmark, not a perfect-information ranking benchmark. \(FN^{ability}\) is the rejection probability among top-quintile latent-ability RAs. \(FN^{merit}\) is the rejection probability among top-quintile realized-merit RAs.
\end{minipage}
\end{table}

Table~\ref{tab:abm_decomposition} separates the channels. Project luck mainly affects the ability-to-merit link: realized merit can diverge from latent ability even when score-ability alignment remains strong. Assessment noise mainly affects the score-to-ability and score-to-merit links. Diagnostic compression weakens the informativeness of routine evidence inside the score. Fixed capacity turns these forms of misalignment into false negatives. Because the idealized row still contains heterogeneous PI environments and a capacity ratio below the top-quintile threshold, it should be read as the model-only benchmark rather than as the minimum feasible false-negative rate. The full noisy-AI environment combines the channels and produces the largest selection-error rates.

\subsection{Results}

\begin{figure}[t]
\centering
\includegraphics[width=\linewidth]{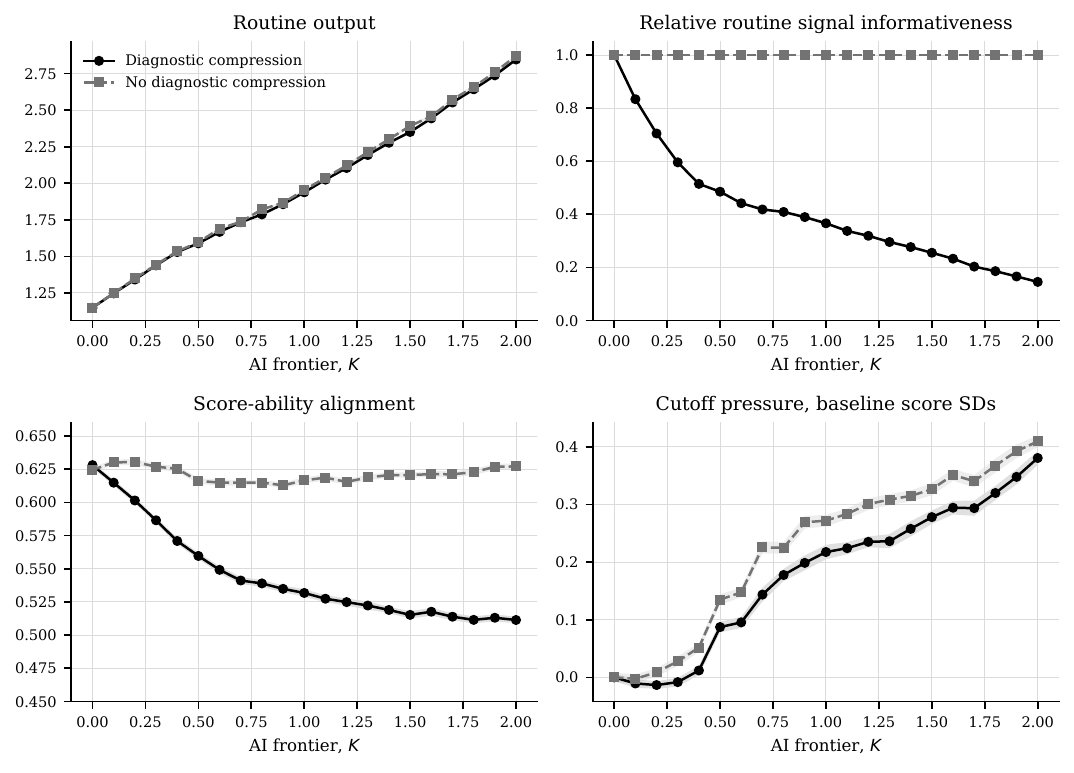}
\caption{Unified simulation mechanisms. The horizontal axis in all panels is the AI frontier \(K\). Solid lines use the diagnostic-compression environment, in which routine-task AI lowers the informativeness of routine evidence; dashed lines use the no-compression counterfactual, which uses the same data-generating distributions and capacity rule while setting \(\rho_R(K_R^T)=1\). Panel A plots mean routine output, showing the production gain from AI. The two lines nearly overlap because diagnostic compression changes how routine evidence is evaluated, not how routine output is produced. Panel B plots routine signal informativeness, \(\rho_R(\alpha_\lambda K)^2/\sigma_R^2\), normalized to one at \(K=0\); a value of \(0.15\) means that routine evidence retains about 15 percent of its baseline informativeness. Panel C plots score-ability alignment, \(\mathrm{corr}(S^{\mathrm{obs}},\theta^L)\), where \(S^{\mathrm{obs}}\) is the perceived admissions score and \(\theta^L\) is latent RA ability. Panel D plots cutoff pressure, \((S^*(K)-S^*(0))/\mathrm{sd}(S^{\mathrm{obs}}_{K=0})\), so the admissions cutoff increase is measured in baseline score standard deviations. Shaded bands are 95 percent Monte Carlo intervals. The figure is illustrative and is not a structural estimate.}
\label{fig:abm_unified_mechanisms_main}
\end{figure}

Fig.~\ref{fig:abm_unified_mechanisms_main} reports the integrated mechanism. Routine output rises as the AI frontier advances. The near overlap between the two lines in Panel A is intentional: diagnostic compression affects the evidentiary loading of routine output, not the production technology itself. To make the evidentiary scale interpretable, Panel B reports routine signal informativeness relative to its \(K=0\) value rather than the raw precision \(\rho_R(\alpha_\lambda K)^2/\sigma_R^2\). Under diagnostic compression, relative routine informativeness falls from \(1.00\) at \(K=0\) to about \(0.15\) at \(K=2\), so routine evidence retains only about 15 percent of its baseline informativeness. Panel C reports score-ability alignment rather than score-merit alignment, because realized merit also contains project luck and novelty-related uncertainty. The alignment remains positive but declines from about \(0.63\) to about \(0.51\). Thus, the simulation does not imply that admissions become random; it shows that evaluated scores become less tightly linked to latent ability. Panel D reports cutoff pressure as the required cutoff increase in \(K=0\) score standard deviations. In the diagnostic-compression environment, this standardized cutoff pressure rises to about \(0.38\) baseline score standard deviations by \(K=2\). The no-compression benchmark produces slightly stronger cutoff pressure because routine-output gains transmit more fully into evaluated scores when routine evidence remains diagnostic. Lower cutoff pressure under compression should not be interpreted as a welfare improvement. It reflects the fact that AI-enabled routine gains are discounted by the evaluation rule; the main cost of compression appears in weaker alignment and higher false negatives, not necessarily in a larger raw cutoff.

The continuous PI-orientation margin is reported in the online appendix. The same payoff function that generates the two-type analytical segmentation also generates monotone policy regions when \(\lambda\) is distributed: lower-\(\lambda\) laboratories choose higher routine-task AI shares and larger team intensity, while higher-\(\lambda\) laboratories preserve more mentoring and less routine-task automation. The plot is placed in the appendix because it is a robustness check for Proposition~\ref{prop:segmentation_revised}, whereas the main simulation result concerns how distributed ability, luck, and noisy evaluation affect selection under fixed capacity. The appendix also reports a 61-point \(\lambda\)-grid check showing that the segmentation pattern is not an artifact of the baseline 31-point grid, and effort/assignment checks showing that the selection-error results do not depend on the baseline effort-loading rule or on filled-random RA assignment.

\begin{figure}[t]
\centering
\includegraphics[width=\linewidth]{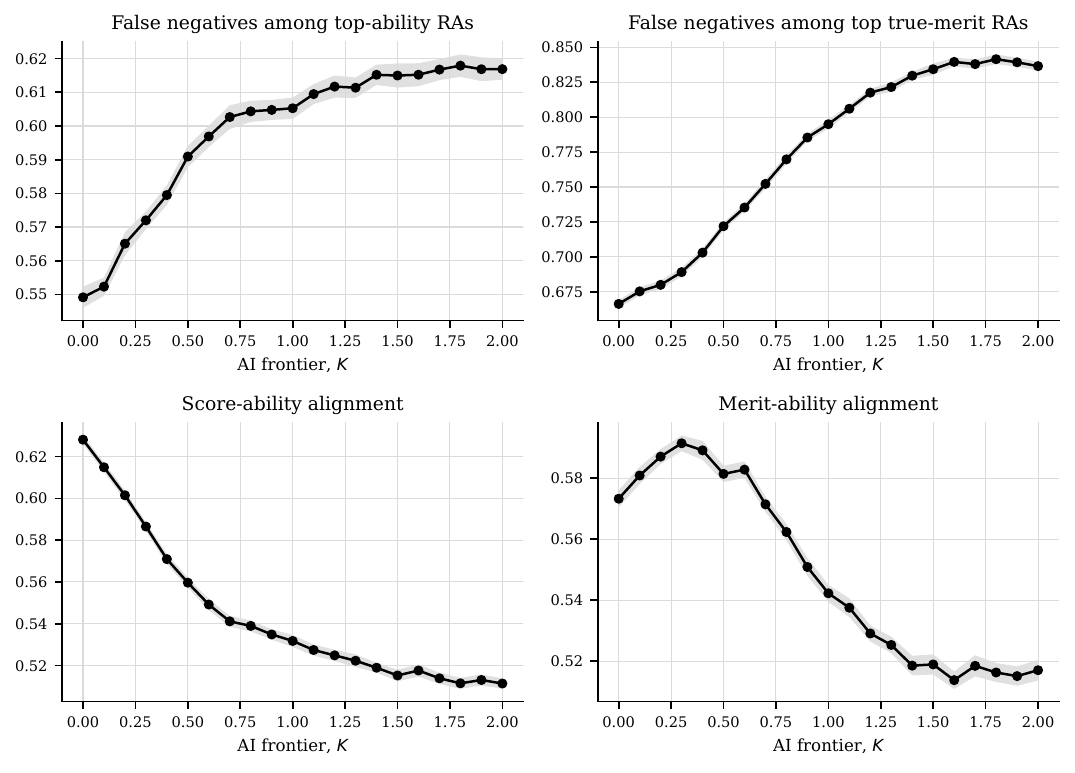}
\caption{Ability, realized merit, and selection errors. All panels use the diagnostic-compression environment and vary the AI frontier \(K\). Admissions are assigned to the top \(Q\) candidates by perceived score \(S^{\mathrm{obs}}\), while ex post merit is measured by \(M^{\mathrm{true}}=Y+\zeta\). Panel A plots \(FN^{ability}\), the probability that a candidate is rejected conditional on latent ability \(\theta_i^L\) being in the top quintile. Panel B plots \(FN^{merit}\), the probability that a candidate is rejected conditional on realized merit \(M_i^{\mathrm{true}}\) being in the top quintile. Because the baseline capacity ratio is \(Q/M=0.17<0.20\), these false-negative rates are mechanically positive even under very accurate ranking. They should be read as simulation-indexed selection-error measures, not calibrated market rejection rates; the informative object is their change as \(K\) rises. Panel C plots score-ability alignment, \(\mathrm{corr}(S^{\mathrm{obs}},\theta^L)\). Panel D plots merit-ability alignment, \(\mathrm{corr}(M^{\mathrm{true}},\theta^L)\), which captures how project luck and PI environments make realized merit diverge from latent ability. Shaded bands are 95 percent Monte Carlo intervals.}
\label{fig:abm_selection_errors_main}
\end{figure}

Fig.~\ref{fig:abm_selection_errors_main} focuses on ability, realized merit, and bottlenecked selection. In the diagnostic-compression environment, the false negative rate among top latent-ability RAs rises from about \(0.55\) at \(K=0\) to about \(0.62\) at \(K=2\). The corresponding false negative rate among top realized-merit RAs rises from about \(0.67\) to about \(0.84\). These levels should not be interpreted as calibrated rejection rates for a real admissions market; the relevant result is the increase relative to the common-capacity benchmark. Score-ability alignment falls from about \(0.63\) to about \(0.51\), while merit-ability alignment falls from about \(0.57\) to about \(0.52\). Thus, AI raises production, but diagnostic compression, project luck, and noisy assessment weaken both links in the selection chain: ability to realized merit, and realized merit to evaluated score.

Additional sensitivity checks in the online appendix separate bottleneck and uncertainty effects. At \(K=2\), increasing the capacity ratio from \(Q/M=0.10\) to \(Q/M=0.35\) lowers the ability-based false negative rate from about \(0.74\) to about \(0.38\), while also reducing the cutoff. Holding \(K=0\) to isolate assessment uncertainty from AI diagnostic compression, higher reception noise raises ability-based false negatives and lowers score-merit alignment. Higher project luck has little direct effect on ability-based false negatives because admissions do not observe \(\zeta\), but it sharply reduces score-merit alignment by widening the gap between latent ability and realized project outcomes.

Overall, the simulation strengthens rather than replaces the propositions. Proposition~\ref{prop:dual_ai_signal} explains why routine evidence can lose diagnostic content. Proposition~\ref{prop:segmentation_revised} explains why organizational choices differ across PI orientations. Proposition~\ref{prop:arms_race_revised} explains why fixed capacity converts broad score improvements into cutoff pressure. The simulation shows that these mechanisms continue to operate when PI orientation and RA talent are distributed, true merit contains project luck, and admissions committees observe noisy perceived scores rather than true quality. 
The simulation uses simple agent-level rules to show how noisy evaluation under scarce capacity can generate aggregate misallocation, even when evaluators do not intentionally misclassify candidates.

\section{Discussion and Policy Implications}
\label{sec:discussion_policy}

The central implication of the model is that generative AI can separate three objects that were previously more closely aligned: what research work produces, what that work reveals about an early-career researcher, and how scarce opportunities are allocated. Pre-doctoral research work is therefore not only project labor. It is also evidence used to evaluate future research potential.

Proposition~\ref{prop:dual_ai_signal} shows that routine-task AI can raise visible routine output while lowering its diagnostic value. This gives an AI-specific reason for a broader concern in responsible research assessment: easily measured outputs should not be treated as direct substitutes for expert judgment \citep{cagan2013dora,hicks2015leiden}.
The implication is not to ignore routine evidence. Routine execution still matters. But when routine artifacts become cheaper and more polished, evaluation needs additional evidence about process and individual contribution. This raises the value of independent process signals: oral research defenses, code and analysis walk-throughs, short replication tasks, research-design interviews, AI-use statements, and offline tests that separate routine execution from independent reasoning.\footnote{
A simple way to formalize this point is to add an independent process signal to the evaluation rule. Suppose the market observes
\[
s_I=\theta+\varepsilon_I,\qquad \varepsilon_I\sim N(0,\sigma_I^2),
\]
where \(s_I\) represents an oral research defense, code walk-through, replication task, or research-design interview. In the linear-Gaussian benchmark, the weight on this independent signal is
\[
\omega_I
=
\frac{1/\sigma_I^2}
{
1/\sigma_\theta^2
+
\rho_R(K_R^T)^2/\sigma_R^2
+
\kappa_\lambda(K_N^T)^2/\sigma_N^2
+
1/\sigma_I^2
}.
\]
Holding the precision of \(s_I\) fixed, a decline in the diagnostic precision of routine evidence, \(\rho_R(K_R^T)^2/\sigma_R^2\), raises the relative weight placed on \(s_I\). Thus, independent process evidence becomes more valuable when routine artifacts lose diagnostic content.
} 
These signals are valuable not only because they are immune to AI assistance, but also because they provide evidence about reasoning, interpretation, and ownership.
The model therefore suggests a shift from evaluating polished artifacts alone toward asking candidates to explain, reproduce, defend, or extend their work.

This shift should be designed carefully. Oral or process-based assessment is not automatically fairer. If implemented informally, it may introduce new biases related to language, confidence, advising access, or institutional background. Its value depends on structured tasks, clear rubrics, and an effort to evaluate reasoning rather than polish or status. This point is especially important because judgment, originality, interpretation, and problem formulation may be harder to assess ex ante than standardized execution \citep{banalestanol2019evaluation,aslan2024unexpectedness}.

Proposition~\ref{prop:segmentation_revised} shows that AI need not make laboratories converge on one organizational form. Some laboratories may use AI to scale routine execution. Others may use AI to support mentoring, research design, and less standardized work. The model does not rank these forms universally. It shows that PI objectives and task technologies can make different AI-enabled organizations privately attractive. Evaluation systems can then reinforce one direction or the other. If they reward visible throughput, they strengthen incentives for scalable routine pipelines. If they reward research design, interpretation, mentoring, and individual contribution, they support apprenticeship-intensive work. This is consistent with evidence that performance-based evaluation systems shape organizational priorities through rankings, prestige, and measurable targets \citep{hicks2012performance}.

The same AI frontier may also create unequal effective capability across laboratories. The baseline treats \(K\) as common to isolate the allocation mechanism, but real laboratories differ in data access, software infrastructure, verification routines, AI expertise, and mentoring capacity. This maps onto absorptive capacity: organizations differ in their ability to recognize, assimilate, and use external knowledge \citep{cohen1990absorptive}. The policy implication is that AI access is not only a software problem. Training, documentation norms, shared infrastructure, and verification support matter if AI is to improve research training rather than widen organizational differences.

Proposition~\ref{prop:arms_race_revised} shows that stronger records are not the same as broader access. When elite admissions capacity is fixed, a broad upward shift in visible records raises the cutoff rather than proportionally expanding opportunity. If AI also draws more candidates into the market, cutoff pressure can rise further. If capacity expands, that pressure is weakened. This follows the rank-order logic of scarce academic tournaments \citep{lazear1981rank,hopkins2012job,hopkins2023is} and is consistent with work on publication bottlenecks and escalating standards in academic careers \citep{card2013nine,ellison2002slowdown}. The implication is not simply that PhD slots should expand. Advising resources, funding, and field-level demand constrain capacity. The more general point is that institutions should monitor whether AI raises the cost of becoming competitive even when it improves absolute preparation.

A related risk is greater reliance on contextual credibility.\footnote{The model captures this only in reduced form through \(r_\lambda\), not as a dynamic reputation process.} If routine artifacts become less informative, evaluators may lean more heavily on PI prestige, institutional affiliation, or prior placement history. Science studies has long emphasized that recognition and resources can cumulate over time \citep{merton1968matthew,bol2018matthew}, and institutional position can matter in scientific resource allocation \citep{viner2004institutionalized}. A better response is to document individual contribution while evaluating research context without reducing it to prestige.

The apprenticeship and welfare extensions qualify the baseline mechanism. Pre-doctoral work may build human capital, not only signals. When AI-supported work is paired with mentoring and exposure to difficult novel tasks, it can raise durable research capability.\footnote{
The online appendix formalizes this channel with a recursive extension in which future research capability increases with effort, mentoring, and exposure to novel-task AI.
} In a learning-dominant regime, AI helps RAs build skills beyond the current admissions cycle. In a signaling-dominant regime, effort mainly improves current admissions scores and may reflect tournament congestion.
The welfare effect is therefore not mechanically positive or negative. AI can raise social value through knowledge output and human-capital formation, but fixed-capacity selection can convert part of the private gain into higher cutoffs and signal escalation as \citet{lazear1981rank} shows. The model identifies when AI-supported work becomes durable capability and when it mainly intensifies competitive signaling.

The model also suggests an empirical agenda. The diagnostic-compression mechanism predicts changes in both evaluation language and evaluation procedures. Recommendation letters and RA evaluations should place less weight on routine execution and more weight on judgment, interpretation, research design, ownership, and initiative after the diffusion of generative AI. Admissions and hiring processes may also add independent process signals, such as oral defenses, code walk-throughs, replication tasks, AI-use statements, and research-design interviews. The segmentation mechanism predicts uneven change across laboratories: AI-assisted pipeline environments should emphasize documentation and verification, while mentoring-intensive environments should emphasize research design and independent reasoning. These predictions can be studied with pre-doctoral job descriptions, recommendation-letter text, admissions procedures, placement outcomes, and measures of laboratory AI infrastructure.

\section{Conclusion}

This paper develops a model of how generative AI changes the pre-doctoral academic labor market. The central mechanism is a wedge between productive value and evidentiary value. AI can make routine research output easier to produce, while making that same output less informative about the RA's own effort, judgment, and research potential. Because pre-doctoral work serves both as research labor and as evidence for PhD admissions, this wedge matters for both laboratory organization and candidate evaluation.

The analytical model yields three main implications. First, routine-task AI can raise observable routine output while reducing the diagnostic precision of routine evidence. Second, heterogeneous PI objectives and task complementarity can generate segmented laboratory strategies: some environments tilt toward scalable routine production, while others rely more on mentoring and novel-task augmentation. Third, when elite admissions capacity is fixed, broad improvements in visible records can raise the admissions cutoff rather than expand access proportionally.

The simulation shows that these mechanisms continue to operate in a richer environment with heterogeneous RAs and PIs, luck in realized research outcomes, noisy admissions evaluation, and fixed-capacity selection. It shows that when ability, realized merit, and evaluated scores can diverge, diagnostic compression and scarce capacity can produce selection errors: high-ability or high-realized-merit candidates may be missed even as AI raises routine output.

The broader lesson is that productivity gains, learning gains, and signaling gains need not coincide. AI may improve research output and help RAs build durable human capital, but it may also intensify competition when stronger visible records are evaluated in a fixed-capacity admissions market. The paper provides a benchmark for studying how AI changes the link between research work, evaluation, and access to scientific careers, and suggests that research-training institutions may need more process-based evidence of individual contribution as routine artifacts become less diagnostic.

\section*{Funding Statement}

This research was sponsored by the Army Research Laboratory and was accomplished under Cooperative Agreement Number W911NF-23-2-0224. The views and conclusions contained in this document are those of the authors and should not be interpreted as representing the official policies, either expressed or implied, of the Army Research Laboratory or the U.S. Government. The U.S. Government is authorized to reproduce and distribute reprints for Government purposes notwithstanding any copyright notation herein.

\clearpage
\processdelayedfloats
\clearpage

\bibliography{references}

\end{document}


{\setstretch{.8}
\maketitle
}

\appendix
\renewcommand{\thetable}{A\arabic{table}}
\renewcommand{\thefigure}{A\arabic{figure}}
\renewcommand{\theequation}{A\arabic{equation}}
\makeatletter
\renewcommand{\thepostfigure}{A\arabic{postfigure}}
\renewcommand{\theposttable}{A\arabic{posttable}}
\renewcommand{\thepostfig}{A\arabic{postfig}}
\renewcommand{\theposttbl}{A\arabic{posttbl}}
\makeatother

\section{Appendix Roadmap and Regularity Conditions}
\label{app:technical}

This online appendix provides the technical and supplementary support for the main text. Sections~\ref{app:truncated_posterior}--\ref{app:proofs} contain the core analytical derivations: signal extraction, equilibrium lemmas, and proofs of the main propositions. Sections~\ref{app:topkis}--H provide additional theoretical material, including an optional route to stronger monotone comparative statics, the dynamic apprenticeship extension, the reduced-form welfare accounting, and the project-risk formulation. Section~\ref{app:simulation_parameters} reports the simulation details, including the parameter table, sensitivity checks, implementation robustness checks, and supplementary simulation figures.

We work with the baseline environment defined in the main text, which combines task-based production, noisy signal extraction, and fixed-capacity tournament assignment \citep{acemoglu2018modeling, daley2014market, lazear1981rank}. Assumptions A1--A6 and A9 define the economic primitives. Assumptions A7, A8, and A10 are regularity conditions used for unique effort choices, simple sorting sets, and local comparative statics. The simulation material in Section~\ref{app:simulation_parameters} uses these primitives as the basis for an illustrative mechanism-preserving exercise; it is not a separate structural calibration.

\subsection*{Economic primitives}

\paragraph{Assumption A1 (Choice sets and filled team intensity).}
For each PI type \(\lambda\in\{\lambda_Q,\lambda_N\}\),
\[
n_\lambda \in [0,\bar n],
\qquad
\alpha_\lambda \in [0,1],
\qquad
g_\lambda \in [0,\bar g],
\]
with \(\bar n,\bar g>0\). The AI capability frontier satisfies \(K\in[0,\bar K]\) and is exogenous. The variable \(n_\lambda\) denotes filled normalized team intensity in equilibrium; posted vacancies and unfilled positions are not separately modeled.

\paragraph{Assumption A2 (Smooth primitives).}
The functions \(a_R(\cdot)\), \(m_R(\cdot)\), \(\sigma_A^2(\cdot)\), \(\kappa_\lambda(\cdot)\), and \(\chi(\cdot,\cdot)\) are continuously differentiable on the feasible domain and twice continuously differentiable around the interior reference points used for local comparative statics. They satisfy
\[
a_R'(\cdot)>0,\quad
m_R(\cdot)>0,\quad
m_R'(\cdot)\ge 0,
\]
\[
\sigma_A^2(\cdot)\ge 0,\quad
(\sigma_A^2)'(\cdot)>0,\quad
\kappa_\lambda(0)=1,\quad
\kappa_\lambda'(\cdot)>0.
\]
The diagnostic loading \(\rho_R(\cdot)\) is derived from Assumption A4 rather than imposed as a separate decreasing primitive.

\paragraph{Assumption A3 (Ability and uncertainty).}
RA ability satisfies
\[
\theta\sim \mathrm{TN}(\mu_\theta,\sigma_\theta^2;\theta_L,\theta_H),
\qquad
0<\theta_L<\theta_H<\infty.
\]
Let \(f\) denote the corresponding density. The noise terms are mutually independent and independent of \(\theta\):
\[
\varepsilon_R\sim N(0,\sigma_R^2),
\qquad
\varepsilon_N\sim N(0,\sigma_N^2),
\qquad
\zeta\sim N(0,\sigma_\zeta^2),
\]
with \(\sigma_R^2,\sigma_N^2,\sigma_\zeta^2>0\).

\paragraph{Assumption A4 (Task outputs and signals).}
For each PI type \(\lambda\),
\[
K_R^T=\alpha K,
\qquad
K_N^T=(1-\alpha)K.
\]
Task outputs are
\[
y_R=a_R(K_R^T)+m_R(K_R^T)(\theta+\eta_R e),
\]
\[
y_N=\kappa_\lambda(K_N^T)(\theta+\eta_N e+\psi g),
\]
where \(\eta_R,\eta_N,\psi>0\).

Routine evidence is summarized by
\[
s_R
=
a_R(K_R^T)+\rho_R(K_R^T)(\theta+\eta_R e)+\varepsilon_R,
\qquad
s_N=y_N+\varepsilon_N.
\]
The diagnostic loading is generated by the artifact-level attribution problem
\[
z_R=h_R+A_R(K_R^T)+\nu_R,
\qquad
h_R=\theta+\eta_R e,
\]
where
\[
A_R(K_R^T)\sim N\!\left(0,\sigma_A^2(K_R^T)\right),
\qquad
\nu_R\sim N(0,\sigma_\nu^2).
\]
With local residual human-input variance \(\sigma_h^2>0\),
\[
\rho_R(K_R^T)
=
\frac{\sigma_h^2}
{\sigma_h^2+\sigma_A^2(K_R^T)+\sigma_\nu^2}.
\]
Since \((\sigma_A^2)'(\cdot)>0\), \(\rho_R(K_R^T)\) is decreasing in routine-task AI. This is an attribution-risk channel: routine-task AI lowers diagnosticity only when it raises the variance of the inseparable non-human component of the artifact. The baseline holds \(\sigma_R^2\) fixed to isolate this loading channel.

\paragraph{Assumption A5 (CES research production).}
Research output for PI type \(\lambda\) is
\[
Y_\lambda
=
\left[
\tau y_R^{\frac{\varsigma_\lambda-1}{\varsigma_\lambda}}
+
(1-\tau)y_N^{\frac{\varsigma_\lambda-1}{\varsigma_\lambda}}
\right]^{\frac{\varsigma_\lambda}{\varsigma_\lambda-1}},
\qquad
\tau\in(0,1),\quad \varsigma_\lambda>0.
\]
For \(\varsigma_\lambda=1\), \(Y_\lambda=y_R^\tau y_N^{1-\tau}\). We allow \(\varsigma_{\lambda_Q}\neq \varsigma_{\lambda_N}\) and maintain \(y_R>0\), \(y_N>0\) on the relevant support.

\paragraph{Assumption A6 (Effort capacity).}
The effort-capacity function \(\chi(\theta,\alpha K)\) is strictly positive and satisfies
\[
\frac{\partial \chi(\theta,\alpha K)}{\partial \theta}>0,
\qquad
\frac{\partial \chi(\theta,\alpha K)}{\partial (\alpha K)}>0.
\]

\paragraph{Assumption A9 (Breakthrough monotonicity).}
For each PI type \(\lambda\), realized novel-project merit is
\[
q_N=y_N+\zeta.
\]
Holding fixed the induced effort schedule and participant distribution, the average breakthrough probability
\[
\bar p_{B,\lambda}=\Pr(q_N>\bar q)
\]
is increasing in novel-task AI \((1-\alpha_\lambda)K\) and mentoring intensity \(g_\lambda\).

\subsection*{Regularity conditions}

\paragraph{Assumption A7 (Effort concavity).}
For each \(\lambda\) and feasible \((\theta,K,\alpha,g,S^*)\),
\[
\frac{1}{\chi(\theta,\alpha K)}
>
\beta_{RA}V
\frac{\bigl(B_\lambda^{B}(K,\alpha,\lambda)\bigr)^2}{(\Sigma_\lambda^{B})^2}
\sup_{z\in\mathbb{R}} |z\varphi(z)|.
\]
This sufficient condition makes the RA effort problem strictly concave.\footnote{
Since \(\sup_{z\in\mathbb R}|z\varphi(z)|=1/\sqrt{2\pi e}\), the condition requires effort-cost curvature to dominate the largest possible curvature contribution from the admissions-probability term.
}
If it fails, effort should be treated as an optimal-effort correspondence rather than a unique function.

\paragraph{Assumption A8 (Sorting).}
For each PI type \(\lambda\), the indirect utility
\[
\widetilde U_\lambda(\theta)
\equiv
U_{RA}\bigl(\theta,e_\lambda^*(\theta)\bigr)
\]
is continuous and strictly increasing in \(\theta\) on the relevant region. The sorting set is
\[
\Theta_\lambda
=
\Bigl\{
\theta\in[\theta_L,\theta_H]:
\widetilde U_\lambda(\theta)\ge \widetilde U_{\lambda'}(\theta)
\text{ for all } \lambda'\neq\lambda,
\quad
\widetilde U_\lambda(\theta)\ge \bar U
\Bigr\}.
\]
We maintain the single-crossing condition under which \(\Theta_\lambda\) has a cutoff representation. The sorting set determines the ability composition of filled positions; it is not an additional participation multiplier on \(n_\lambda\).

\paragraph{Assumption A10 (Local PI regularity).}
For each PI type and cutoff \(S^*\), the PI objective is continuous on the compact choice set and attains a nonempty set of maximizers. Around the interior reference points used for local comparative statics, the maximizer is isolated, the objective is twice continuously differentiable, and the Hessian with respect to \((n_\lambda,\alpha_\lambda,g_\lambda)\) is nonsingular. The comparative statics in the main text are local, not global monotonicity claims.

\section{Signal Extraction: Exact Truncated Updating and the Linear Approximation}
\label{app:truncated_posterior}

The main text uses a linear prediction rule to keep the evaluation mechanism transparent. This section clarifies the relationship between that rule and exact Bayesian updating when ability has bounded support. The derivation is a residualized benchmark: it is used to obtain the signal weights, not to assume that admissions committees directly observe effort.

The diagnostic loading \(\rho_R(K_R^T)\) is derived in Assumption~A4 from an artifact-level attribution problem. A larger routine-task AI allocation raises the variance of the inseparable non-human component of the routine artifact, so the projection loading of routine evidence on human input declines. The signal-extraction algebra below treats this derived loading as the routine-signal coefficient.

Define the residualized signal vector
\[
\bar{\mathbf s}
=
\begin{pmatrix}
\bar s_R\\
\bar s_N
\end{pmatrix}
=
H_\lambda(\theta-\mu_\theta)+\varepsilon,
\qquad
H_\lambda
=
\begin{pmatrix}
\rho_R(K_R^T)\\
\kappa_\lambda(K_N^T)
\end{pmatrix},
\]
where
\[
\varepsilon\sim N(0,\Sigma_\varepsilon),
\qquad
\Sigma_\varepsilon=\mathrm{diag}(\sigma_R^2,\sigma_N^2),
\]
and
\[
\bar s_R
=
\tilde s_R-\rho_R(K_R^T)\eta_R e,
\qquad
\bar s_N
=
\tilde s_N-\kappa_\lambda(K_N^T)(\eta_N e+\psi g).
\]
This residualization is only a technical device for deriving the benchmark weights. In the main text, the same weights are used in the reduced-form linear evaluation rule from noisy task-level evidence.

If the prior were unbounded Gaussian,
\[
\theta\sim N(\mu_\theta,\sigma_\theta^2),
\]
standard normal updating would give posterior variance
\begin{equation}
V_\lambda
=
\left[
\frac{1}{\sigma_\theta^2}
+
H_\lambda'\Sigma_\varepsilon^{-1}H_\lambda
\right]^{-1},
\label{eq:appendix_unbounded_post_var}
\end{equation}
and posterior mean
\begin{equation}
m_\lambda(\bar{\mathbf s})
=
\mu_\theta
+
V_\lambda H_\lambda'\Sigma_\varepsilon^{-1}\bar{\mathbf s}.
\label{eq:appendix_unbounded_post_mean}
\end{equation}
Equation~\eqref{eq:appendix_unbounded_post_mean} is the source of the linear prediction weights used in the main text.

With bounded support,
\[
\theta\sim \mathrm{TN}(\mu_\theta,\sigma_\theta^2;\theta_L,\theta_H),
\]
the posterior is the same Gaussian kernel restricted to \([\theta_L,\theta_H]\):
\[
\theta\mid \bar{\mathbf s},\ \theta\in[\theta_L,\theta_H]
\sim
\mathrm{TN}\bigl(m_\lambda(\bar{\mathbf s}),V_\lambda;\theta_L,\theta_H\bigr).
\]
Using the standard first moment of a truncated normal distribution \citep{tallis1961moment,horrace2005some}, define
\[
a_\lambda(\bar{\mathbf s})
=
\frac{\theta_L-m_\lambda(\bar{\mathbf s})}{\sqrt{V_\lambda}},
\qquad
b_\lambda(\bar{\mathbf s})
=
\frac{\theta_H-m_\lambda(\bar{\mathbf s})}{\sqrt{V_\lambda}}.
\]
The exact posterior mean is
\begin{equation}
\mathbb{E}[\theta\mid \bar{\mathbf s},\theta\in[\theta_L,\theta_H]]
=
m_\lambda(\bar{\mathbf s})
+
\sqrt{V_\lambda}
\frac{
\varphi\!\left(a_\lambda(\bar{\mathbf s})\right)
-
\varphi\!\left(b_\lambda(\bar{\mathbf s})\right)
}{
\Phi\!\left(b_\lambda(\bar{\mathbf s})\right)
-
\Phi\!\left(a_\lambda(\bar{\mathbf s})\right)
}.
\label{eq:appendix_exact_truncated_posterior}
\end{equation}

The truncation correction in \eqref{eq:appendix_exact_truncated_posterior} makes the exact posterior mean nonlinear near the support boundaries. This is why the main text treats the linear formula as an exact rule only in the unbounded residualized Gaussian benchmark, and otherwise as a tractable linear predictor from noisy task-level evidence.

The key informativeness result is unchanged. If
\[
\rho_R(K_R^T)\to0,
\]
then the routine component of \(H_\lambda\) converges to zero. Consequently, both \(V_\lambda\) and \(m_\lambda(\bar{\mathbf s})\) become independent of \(\bar s_R\). The exact truncated posterior mean in \eqref{eq:appendix_exact_truncated_posterior} also becomes independent of \(\bar s_R\). Thus, even under exact truncated-normal updating, routine evidence does not update beliefs about ability when it carries no diagnostic loading.

\section{Auxiliary Lemmas for the Baseline Equilibrium}
\label{app:auxiliary_lemmas}

This section collects the auxiliary results used to construct the baseline equilibrium. 
Lemma~\ref{lem:bayes_weights_appendix} records the benchmark signal weights. 
Lemma~\ref{lem:reduced_form_score_appendix} rewrites the admissions score in reduced form. 
Lemma~\ref{lem:effort_unique_appendix} gives uniqueness of RA effort. 
Lemma~\ref{lem:participation_cutoff_appendix} states the sorting representation. 
Proposition~\ref{prop:existence_appendix} gives the local market-clearing cutoff.

\begin{lemma}[Benchmark linear prediction weights]
\label{lem:bayes_weights_appendix}
Let \(K_R^T=\alpha K\) and \(K_N^T=(1-\alpha)K\). Define the centered signals
\[
\tilde s_R=s_R-a_R(K_R^T)-\rho_R(K_R^T)\mu_\theta,
\qquad
\tilde s_N=s_N-\kappa_\lambda(K_N^T)\mu_\theta.
\]
The benchmark linear evaluation rule is
\[
\hat\theta
=
\mu_\theta
+
\omega_R^{B}(\alpha,K,\lambda)\tilde s_R
+
\omega_N^{B}(\alpha,K,\lambda)\tilde s_N,
\]
where
\begin{align}
\omega_R^{B}(\alpha,K,\lambda)
&=
\frac{\rho_R(K_R^T)/\sigma_R^2}
{
\sigma_\theta^{-2}
+
\rho_R(K_R^T)^2/\sigma_R^2
+
\kappa_\lambda(K_N^T)^2/\sigma_N^2
},
\label{eq:appendix_omegaR}
\\
\omega_N^{B}(\alpha,K,\lambda)
&=
\frac{\kappa_\lambda(K_N^T)/\sigma_N^2}
{
\sigma_\theta^{-2}
+
\rho_R(K_R^T)^2/\sigma_R^2
+
\kappa_\lambda(K_N^T)^2/\sigma_N^2
}.
\label{eq:appendix_omegaN}
\end{align}
These weights are exact posterior weights in the unbounded residualized Gaussian benchmark and are used as best-linear-prediction weights in the main text.
\end{lemma}

\begin{proof}
The result follows from the residualized Gaussian system in Section~\ref{app:truncated_posterior}. Substituting
\[
H_\lambda
=
\begin{pmatrix}
\rho_R(K_R^T)\\
\kappa_\lambda(K_N^T)
\end{pmatrix},
\qquad
\Sigma_\varepsilon=\mathrm{diag}(\sigma_R^2,\sigma_N^2),
\]
into the normal-normal posterior mean formula gives \eqref{eq:appendix_omegaR}--\eqref{eq:appendix_omegaN}. Reintroducing deterministic effort and mentoring terms changes the intercept and effort loading, but not the benchmark precision weights.
\end{proof}

\begin{lemma}[Reduced-form admissions score]
\label{lem:reduced_form_score_appendix}
Using the benchmark weights and the PI credibility multiplier, the admissions score can be written as
\[
S
=
A_\lambda^{B}(\theta;K,\alpha,g)
+
B_\lambda^{B}(K,\alpha,\lambda)e
+
\Sigma_\lambda^{B}(K,\alpha,\lambda)\xi,
\qquad
\xi\sim N(0,1),
\]
where all weights are evaluated at \((\alpha,K,\lambda)\), and
\begin{align}
A_\lambda^{B}(\theta;K,\alpha,g)
&=
r_\lambda
\Big[
\mu_\theta
+
\big(
\omega_R^{B}\rho_R(K_R^T)
+
\omega_N^{B}\kappa_\lambda(K_N^T)
\big)(\theta-\mu_\theta)
+
\omega_N^{B}\kappa_\lambda(K_N^T)\psi g
\Big],
\label{eq:A_Bayes}
\\
B_\lambda^{B}(K,\alpha,\lambda)
&=
r_\lambda
\Big[
\omega_R^{B}\rho_R(K_R^T)\eta_R
+
\omega_N^{B}\kappa_\lambda(K_N^T)\eta_N
\Big],
\label{eq:B_Bayes}
\\
\Sigma_\lambda^{B}(K,\alpha,\lambda)
&=
r_\lambda
\sqrt{
(\omega_R^{B})^2\sigma_R^2
+
(\omega_N^{B})^2\sigma_N^2
}.
\label{eq:Sigma_Bayes}
\end{align}
Here \(A_\lambda^B\) is the deterministic score component conditional on ability \(\theta\), \(B_\lambda^B\) is the effort loading, and \(\Sigma_\lambda^B\) is the residual score standard deviation. Under the maintained primitives, \(B_\lambda^B(K,\alpha,\lambda)>0\).
\end{lemma}

\begin{proof}
Substitute the centered signals into \(S=r_\lambda\hat\theta\) and collect the terms multiplying \(\theta-\mu_\theta\), effort \(e\), mentoring \(g\), and the Gaussian innovations \((\varepsilon_R,\varepsilon_N)\). Under the maintained primitives, the positive signal loadings and positive noise variances imply \(\omega_R^B,\omega_N^B>0\). Since \(r_\lambda,\rho_R,\kappa_\lambda,\eta_R,\eta_N>0\), it follows that \(B_\lambda^B>0\).
\end{proof}

\begin{lemma}[Unique optimal effort]
\label{lem:effort_unique_appendix}
Under Assumptions A2--A7, for each \((\lambda,\theta,K,\alpha,g,S^*)\), the RA effort problem has a unique optimal solution \(e_\lambda^*(\theta)\). The solution may be interior or at the boundary \(e=0\).
\end{lemma}

\begin{proof}
The RA objective is
\[
U_{RA}(e)
=
\bar w
-
\frac{e^2}{2\chi(\theta,\alpha K)}
+
\beta_{RA}V
\left[
1-\Phi\!\left(
\frac{S^*-A_\lambda^{B}(\theta;K,\alpha,g)-B_\lambda^{B}(K,\alpha,\lambda)e}{\Sigma_\lambda^{B}}
\right)
\right].
\]
It is continuous in \(e\). Since the admission probability is bounded and \(\chi(\theta,\alpha K)>0\), the quadratic effort cost implies \(U_{RA}(e)\to-\infty\) as \(e\to\infty\). Hence a maximizer exists on \([0,\infty)\).

Let
\[
z(e)=
\frac{S^*-A_\lambda^{B}(\theta;K,\alpha,g)-B_\lambda^{B}(K,\alpha,\lambda)e}
{\Sigma_\lambda^{B}}.
\]
Then
\[
\frac{\partial^2 U_{RA}}{\partial e^2}
=
-\frac{1}{\chi(\theta,\alpha K)}
+
\beta_{RA}V
\frac{\big(B_\lambda^{B}(K,\alpha,\lambda)\big)^2}{(\Sigma_\lambda^{B})^2}
z(e)\varphi(z(e)).
\]
Thus,
\[
\frac{\partial^2 U_{RA}}{\partial e^2}
\le
-\frac{1}{\chi(\theta,\alpha K)}
+
\beta_{RA}V
\frac{\big(B_\lambda^{B}(K,\alpha,\lambda)\big)^2}{(\Sigma_\lambda^{B})^2}
\sup_{z\in\mathbb R}|z\varphi(z)|.
\]
Assumption A7 makes this upper bound strictly negative. Therefore \(U_{RA}\) is strictly concave in \(e\), and the maximizer is unique.
\end{proof}

\begin{lemma}[Participation and sorting]
\label{lem:participation_cutoff_appendix}
Under Assumption A8, the sorting set \(\Theta_\lambda\) is well defined. If the maintained single-crossing condition holds on the relevant region, then \(\Theta_\lambda\) admits the stated cutoff representation.
\end{lemma}

\begin{proof}
By Lemma~\ref{lem:effort_unique_appendix}, the indirect utility
\[
\widetilde U_\lambda(\theta)
\equiv
U_{RA}\bigl(\theta,e_\lambda^*(\theta)\bigr)
\]
is well defined. Since \(\widetilde U_\lambda(\theta)\) is continuous and strictly increasing, the participation set
\[
\{\theta\in[\theta_L,\theta_H]:\widetilde U_\lambda(\theta)\ge \bar U\}
\]
is empty, the full support, or a closed upper interval.

When RAs choose across PI segments, monotonicity of \(\widetilde U_\lambda\) alone is not enough to guarantee a cutoff sorting set. We therefore use the maintained single-crossing condition in Assumption A8: for each pair \((\lambda,\lambda')\), the relevant weak-preference set
\[
\{\theta:\widetilde U_\lambda(\theta)\ge \widetilde U_{\lambda'}(\theta)\}
\]
is an interval on the region of interest. Under this maintained single-crossing condition, intersecting the weak-preference set with the participation set gives the interval representation for \(\Theta_\lambda\); in the maintained region this interval is summarized by the cutoff used in the main text.
\end{proof}

\begin{appproposition}[Local market-clearing cutoff]
\label{prop:existence_appendix}
Under Assumptions A1--A10, fix a maintained local region in which PI policies, RA effort schedules, and sorting sets vary continuously with \(S^*\). Suppose the induced aggregate admissions function is continuous and crosses \(Q\) on the relevant cutoff interval. Then a market-clearing cutoff exists. If the aggregate admissions function is strictly decreasing on that interval, the cutoff is locally unique.
\end{appproposition}

\begin{proof}
For each candidate cutoff \(S^*\), Lemmas~\ref{lem:effort_unique_appendix} and~\ref{lem:participation_cutoff_appendix} give an induced effort schedule and sorting set. By compactness and continuity, each type-specific PI problem has a maximizer. In the maintained interior region, Assumption A10 selects an isolated regular maximizer, so the induced policies
\[
(n_\lambda^*(S^*),\alpha_\lambda^*(S^*),g_\lambda^*(S^*))
\]
vary continuously with \(S^*\).

Define aggregate expected admissions by
\[
\Psi(S^*)
=
\sum_{\lambda\in\{\lambda_Q,\lambda_N\}}
\mu_\lambda n_\lambda^*(S^*)
\int
P_\lambda\bigl(\theta,e_\lambda^*(\theta);S^*\bigr)\,dF_\lambda(\theta),
\]
where \(F_\lambda\) is the conditional distribution among RAs filling positions in segment \(\lambda\). Since \(n_\lambda^*(S^*)\) is filled team intensity, no additional participation-rate multiplier appears. If \(n_\lambda\) were modeled as posted capacity, the mass term would instead include a fill rate.

By the maintained local market-clearing condition, \(\Psi(S^*)\) is continuous and crosses \(Q\) on the relevant cutoff interval. The intermediate value theorem gives a solution to
\[
\Psi(S^*)=Q.
\]
If \(\Psi(S^*)\) is strictly decreasing on that interval, this solution is locally unique.
\end{proof}

\section{Proofs of Main Propositions}
\label{app:proofs}

\subsection{Proof of Proposition 1}

\begin{proof}
Let \(x\equiv \alpha K\). The proposition is a partial comparative static holding total AI capability \(K\) and the local human-input term \(h_R=\theta+\eta_Re\) fixed.

For part (i), routine output is
\[
y_R(x)=a_R(x)+m_R(x)h_R.
\]
Thus
\[
\frac{dy_R(x)}{dx}
=
a_R'(x)+m_R'(x)h_R
\ge 0,
\]
because \(a_R'(x)>0\), \(m_R'(x)\ge0\), and \(h_R>0\).

For part (ii), Assumption A4 gives
\[
\rho_R(x)
=
\frac{\sigma_h^2}
{\sigma_h^2+\sigma_A^2(x)+\sigma_\nu^2}.
\]
Since \(d\sigma_A^2(x)/dx>0\),
\[
\rho_R'(x)
=
-
\frac{
\sigma_h^2\,d\sigma_A^2(x)/dx
}
{
\left[\sigma_h^2+\sigma_A^2(x)+\sigma_\nu^2\right]^2
}
<0.
\]
The diagnostic precision of routine evidence about human contribution is
\[
\frac{\rho_R(x)^2}{\sigma_R^2}.
\]
Therefore
\[
\frac{d}{dx}
\left(
\frac{\rho_R(x)^2}{\sigma_R^2}
\right)
=
\frac{2\rho_R(x)\rho_R'(x)}{\sigma_R^2}
<0,
\]
because \(\rho_R(x)>0\) and \(\sigma_R^2>0\).

For part (iii),
\[
\omega_R^{B}(\alpha,K,\lambda)
=
\frac{\rho_R(\alpha K)/\sigma_R^2}
{
\sigma_\theta^{-2}
+
\rho_R(\alpha K)^2/\sigma_R^2
+
\kappa_\lambda((1-\alpha)K)^2/\sigma_N^2
}.
\]
As \(\rho_R(\alpha K)\to0\), the numerator converges to zero while the denominator remains strictly positive. Hence \(\omega_R^B(\alpha,K,\lambda)\to0\). Thus routine-task AI can raise routine output while reducing the usefulness of routine evidence for evaluating human contribution and, in the residualized benchmark, underlying ability.
\end{proof}

\subsection{Proof of Proposition 2}

\begin{proof}
Fix the common interior comparison point
\[
\bar x=(\bar n,\bar\alpha,\bar g),
\]
and evaluate all local derivatives at \(\bar x\), under the local-envelope convention used in the proposition.

Define
\[
D_{\alpha,\lambda}^Y
\equiv
\left.\frac{\partial \bar Y_\lambda}{\partial \alpha_\lambda}\right|_{\bar x},
\qquad
D_{g,\lambda}^Y
\equiv
\left.\frac{\partial \bar Y_\lambda}{\partial g_\lambda}\right|_{\bar x}.
\]
Maintain \(\bar p_{B,Q}\in(0,1)\) at the comparison point. For the quality-oriented breakthrough channel, define
\[
B_\alpha
\equiv
-
\left.\frac{\partial \bar p_{B,Q}}{\partial \alpha_Q}\right|_{\bar x},
\qquad
B_g
\equiv
\left.\frac{\partial \bar p_{B,Q}}{\partial g_Q}\right|_{\bar x}.
\]
Since \(K_N^T=(1-\alpha)K\) and breakthrough probability is increasing in novel-task AI and mentoring, \(B_\alpha>0\) and \(B_g>0\). Let
\[
\Lambda_Q
\equiv
\bar n(1-\bar p_{B,Q})^{\bar n-1},
\qquad
\Delta_Q
\equiv
(1-\bar p_{B,Q})^{\bar n}
\ln\!\left(\frac{1}{1-\bar p_{B,Q}}\right),
\]
and
\[
C_n(\bar x)
\equiv
\bar w+\frac{c_g}{2}\bar g^2+c_n\bar n.
\]
Here \(\Lambda_Q>0\) and \(\Delta_Q>0\) whenever \(\bar p_{B,Q}\in(0,1)\).

The relevant marginal payoff terms are
\[
M_{\alpha,N}
=
\gamma_N\bar nD_{\alpha,N}^Y,
\qquad
M_{\alpha,Q}
=
-\Omega\Lambda_QB_\alpha
+
\gamma_Q\bar nD_{\alpha,Q}^Y,
\]
\[
M_{g,N}
=
\gamma_N\bar nD_{g,N}^Y
-
c_g\bar g\bar n,
\qquad
M_{g,Q}
=
\Omega\Lambda_QB_g
+
\gamma_Q\bar nD_{g,Q}^Y
-
c_g\bar g\bar n,
\]
and
\[
M_{n,N}
=
\gamma_N\bar Y_N-C_n(\bar x),
\qquad
M_{n,Q}
=
\Omega\Delta_Q+\gamma_Q\bar Y_Q-C_n(\bar x).
\]

A sufficient set of primitive thresholds is:
\[
D_{\alpha,N}^Y>0,
\]
\[
\Omega>
\bar\Omega_\alpha(\bar x)
\equiv
\frac{\gamma_Q\bar n[D_{\alpha,Q}^Y]_+}{\Lambda_QB_\alpha},
\qquad
\Omega>
\bar\Omega_g(\bar x)
\equiv
\frac{[c_g\bar g\bar n-\gamma_Q\bar nD_{g,Q}^Y]_+}{\Lambda_QB_g},
\]
\[
\gamma_N
<
\bar\gamma_g(\bar x)
\equiv
\begin{cases}
c_g\bar g/D_{g,N}^Y, & D_{g,N}^Y>0,\\
+\infty, & D_{g,N}^Y\le 0,
\end{cases}
\]
and
\[
\gamma_N
\ge
\underline\gamma_n(\bar x)
\equiv
\frac{C_n(\bar x)}{\bar Y_N},
\qquad
\Omega
\le
\bar\Omega_n(\bar x)
\equiv
\frac{C_n(\bar x)-\gamma_Q\bar Y_Q}{\Delta_Q},
\]
with
\[
C_n(\bar x)>\gamma_Q\bar Y_Q.
\]
These thresholds are jointly feasible if
\[
\max\{\bar\Omega_\alpha(\bar x),\bar\Omega_g(\bar x)\}
<
\bar\Omega_n(\bar x)
\]
and
\[
\underline\gamma_n(\bar x)<\bar\gamma_g(\bar x).
\]

Under these conditions,
\[
M_{\alpha,N}>0>M_{\alpha,Q},
\qquad
M_{g,Q}>0>M_{g,N},
\qquad
M_{n,N}\ge0\ge M_{n,Q}.
\]
The local regularity condition in main-text Proposition 2 implies that these own marginal payoff signs determine the local direction of the corresponding type-specific optima. Therefore, in a local equilibrium neighborhood,
\[
\alpha_N^*>\alpha_Q^*,
\qquad
g_Q^*>g_N^*,
\qquad
n_N^*\ge n_Q^*.
\]
If \(M_{n,N}>0>M_{n,Q}\), then \(n_N^*>n_Q^*\).
\end{proof}

\subsection{Proof of Proposition 3}

\begin{proof}
Under the filled-team-intensity convention, total candidate mass is
\[
M(\bar K)
=
\sum_{\lambda\in\{\lambda_Q,\lambda_N\}}
\mu_\lambda n_\lambda^*(\bar K).
\]
If \(n_\lambda\) were modeled as posted capacity, \(n_\lambda^*(\bar K)\) would be replaced by \(n_\lambda^*(\bar K)\pi_\lambda^*(\bar K)\).

With fixed capacity \(Q\), the cutoff satisfies
\[
1-H_{\bar K}(S^*)=\frac{Q}{M(\bar K)}.
\]
Under the location-shift representation,
\[
S_i(\bar K)=\Delta(\bar K)+u_i,
\qquad
\Delta'(\bar K)>0,
\]
where \(u_i\sim H_0\), we have
\[
H_{\bar K}(s)=H_0(s-\Delta(\bar K)).
\]
Thus
\[
S^*(\bar K)
=
\Delta(\bar K)
+
H_0^{-1}\!\left(1-\frac{Q}{M(\bar K)}\right).
\]
Let
\[
q(\bar K)
\equiv
H_0^{-1}\!\left(1-\frac{Q}{M(\bar K)}\right).
\]
If \(h_0(q(\bar K))>0\), then for fixed \(Q\),
\[
\frac{dS^*(\bar K)}{d\bar K}
=
\Delta'(\bar K)
+
\frac{QM'(\bar K)}
{M(\bar K)^2h_0(q(\bar K))}.
\]
Hence the cutoff rises one-for-one with the common score shift when candidate mass is fixed, and candidate entry strengthens the cutoff increase when \(M'(\bar K)\ge0\).

If capacity also varies with \(\bar K\), redefine
\[
q_{\mathrm{cap}}(\bar K)
\equiv
H_0^{-1}\!\left(1-\frac{Q(\bar K)}{M(\bar K)}\right).
\]
The same calculation gives
\[
\frac{dS^*(\bar K)}{d\bar K}
=
\Delta'(\bar K)
+
\frac{
Q(\bar K)M'(\bar K)/M(\bar K)^2
-
Q'(\bar K)/M(\bar K)
}
{h_0(q_{\mathrm{cap}}(\bar K))}.
\]
Thus capacity expansion weakens the cutoff effect and can offset it if sufficiently large. This last expression is reported only to show how capacity expansion would alter the fixed-capacity benchmark.

Now fix a candidate \((\theta,e)\) in PI segment \(\lambda\). Let
\[
z_\lambda
=
\frac{
S^*
-
A_\lambda^B(\theta;K,\alpha_\lambda,g_\lambda)
-
B_\lambda^B(K,\alpha_\lambda,\lambda)e
}
{\Sigma_\lambda^B}.
\]
Since \(P_\lambda(\theta,e;S^*)=1-\Phi(z_\lambda)\), differentiating while holding \((\theta,e)\) fixed gives
\[
\frac{dP_\lambda}{d\bar K}
=
\frac{\varphi(z_\lambda)}{\Sigma_\lambda^B}
\left[
\frac{dA_\lambda^B}{d\bar K}
+
e\frac{dB_\lambda^B}{d\bar K}
-
\frac{dS^*}{d\bar K}
+
z_\lambda\frac{d\Sigma_\lambda^B}{d\bar K}
\right].
\]
Thus, if the candidate's own score components, including the residual-noise adjustment, do not rise enough to offset the cutoff increase, then
\[
\frac{dP_\lambda}{d\bar K}<0.
\]
If equilibrium effort also varies with \(\bar K\), the same expression includes the additional own-score term
\[
B_\lambda^B(K,\alpha_\lambda,\lambda)\frac{de}{d\bar K}.
\]
The congestion conclusion is unchanged: admission probability falls whenever the candidate's own score response is smaller than the cutoff response.
\end{proof}

\section{A Stronger Optional Monotone-Comparative-Statics Route}
\label{app:topkis}

The purpose of this section is only to clarify what would be needed to obtain a stronger global monotone-comparative-statics result. The main text states Proposition 2 as a local characterization because the baseline model has several interacting margins: filled team intensity, AI allocation, mentoring, RA effort responses, and admissions-cutoff feedback.

A standard route would use lattice-theoretic monotone comparative statics. Let
\[
x_\lambda=(n_\lambda,\alpha_\lambda,g_\lambda)
\]
denote the PI's organizational choice, and let \(t\) be a primitive parameter that raises the return to one organizational margin, such as scalable output, breakthrough value, or mentoring effectiveness. Suppose the feasible choice set \(X\) is a compact sublattice, \(\Pi_\lambda(x_\lambda;t)\) is supermodular in \(x_\lambda\), and \(\Pi_\lambda\) has increasing differences in \((x_\lambda,t)\). Then the set of maximizers is nondecreasing in \(t\) in the strong set order; if the maximizer is unique, the optimal choice \(x_\lambda^*(t)\) is monotone in \(t\) \citep{topkis1978minimizing, topkis1998supermodularity, milgrom1994monotone}.

Applied to the present setting, this approach would require stronger primitive restrictions than the baseline imposes. For example, a parameter that raises the return to scalable output would need to have increasing differences with routine-task AI allocation and team intensity, while a parameter that raises the value of upper-tail research or mentoring would need to have increasing differences with mentoring and novel-task preservation. In derivative notation, sufficient local analogues would include terms such as
\[
\frac{\partial^2 \Pi_N}{\partial \alpha\,\partial t_{\alpha,N}}>0,
\qquad
\frac{\partial^2 \Pi_Q}{\partial g\,\partial t_{g,Q}}>0,
\qquad
\frac{\partial^2 \Pi_N}{\partial n\,\partial t_{n,N}}>0,
\]
together with the relevant supermodularity and single-crossing conditions across the choice variables.

The paper does not impose these global restrictions because they would substantially narrow the model. The local threshold characterization in main-text Proposition 2 is sufficient for the segmentation mechanism emphasized in the main text.

\section{Dynamic Apprenticeship Extension}
\label{app:bellman}

The baseline model focuses on evaluation and congestion. This extension shows how the same environment can also include apprenticeship learning. The purpose is not to solve a full dynamic PI--RA--admissions game, but to isolate how effort, mentoring, and exposure to novel-task AI can build future research capability in a recursive setting \citep{bellman1957dynamic,stokey1989recursive}.

Let \(q_t\) denote the RA's current research capability. At date \(t\), the aggregate state is
\[
x_t=(r_{\lambda t},K_t,S_t^*),
\]
where \(r_{\lambda t}\) is PI credibility, \(K_t\) is the AI capability frontier, and \(S_t^*\) is the admissions cutoff. A type-\(\lambda\) PI has chosen the laboratory environment \((\alpha_{\lambda t},g_{\lambda t})\), and the RA chooses effort \(e_t\). Future capability evolves according to
\begin{equation}
q_{t+1}
=
q_t
+
\ell(e_t,K_{Nt}^T)
+
\psi_L\log(1+g_{\lambda t})
+
\nu_{q,t+1},
\qquad
K_{Nt}^T=(1-\alpha_{\lambda t})K_t,
\label{eq:theta_prime_appendix}
\end{equation}
where \(\ell_e\ge 0\), \(\ell_{K_N}\ge 0\), and \(\nu_{q,t+1}\) is an idiosyncratic learning shock. The law of motion captures learning-by-doing, exposure to novel-task AI, and mentoring.

Let \(V^A(q)\) denote the continuation value after admission to a PhD program, and let \(V^O(q)\) denote the continuation value outside that placement outcome. Write
\[
\mathcal E_{\lambda t}
\equiv
(x_t,\alpha_{\lambda t},g_{\lambda t})
\]
for the current laboratory and evaluation environment. Conditional on this environment, the RA's recursive effort problem is
\begin{equation}
\begin{aligned}
V_{RA,\lambda}(q_t;\mathcal E_{\lambda t})
=
\max_{e_t\ge0}
\Bigg\{
&
\bar w
-
\frac{e_t^2}{2\chi(q_t,\alpha_{\lambda t}K_t)}
\\
&+
\beta_{RA}
\mathbb{E}_t
\left[
P_{\lambda t}(q_t,e_t;S_t^*)V^A(q_{t+1})
+
\bigl(1-P_{\lambda t}(q_t,e_t;S_t^*)\bigr)V^O(q_{t+1})
\right]
\Bigg\}.
\end{aligned}
\label{eq:ra_bellman_appendix}
\end{equation}

This problem separates two returns to effort. The current signaling return operates through the admission probability \(P_{\lambda t}(q_t,e_t;S_t^*)\). The learning return operates through future capability \(q_{t+1}\). To see this, suppose the objective is differentiable and the optimum is interior. Let
\[
P_t
\equiv
P_{\lambda t}(q_t,e_t;S_t^*),
\qquad
P_e(q_t,e_t;S_t^*)
\equiv
\frac{\partial P_{\lambda t}(q_t,e_t;S_t^*)}{\partial e_t}.
\]
The first-order condition can be written as
\begin{equation}
\begin{aligned}
\frac{e_t}{\chi(q_t,\alpha_{\lambda t}K_t)}
=&\;
\underbrace{
\beta_{RA}
\mathbb{E}_t
\left[
P_e(q_t,e_t;S_t^*)
\bigl(V^A(q_{t+1})-V^O(q_{t+1})\bigr)
\right]
}_{\text{signaling return}}
\\
&+
\underbrace{
\beta_{RA}
\mathbb{E}_t
\left[
\bigl(P_tV^{A\prime}(q_{t+1})+(1-P_t)V^{O\prime}(q_{t+1})\bigr)
\ell_e(e_t,K_{Nt}^T)
\right]
}_{\text{learning return}}.
\end{aligned}
\label{eq:effort_foc_learning_appendix}
\end{equation}
The first term is the signaling return to effort: effort raises the probability of admission. The second term is the learning return: effort raises future capability, which is valuable when continuation values increase in \(q_{t+1}\).

If \(V^A(q)\) and \(V^O(q)\) are constant in \(q\), the learning-return term disappears and the problem collapses to the static signaling problem in the main text. If continuation values are increasing in \(q\), effort has an additional human-capital return. Mentoring has the same apprenticeship logic. From \eqref{eq:theta_prime_appendix},
\[
\frac{\partial \mathbb{E}_t[q_{t+1}]}{\partial g_{\lambda t}}
=
\frac{\psi_L}{1+g_{\lambda t}}>0.
\]
Thus, when future capability is valuable, mentoring and novel-task exposure can build durable research capability rather than merely improve current admissions signals.

A full dynamic PI problem would require additional transition laws for PI credibility, the AI frontier, and future admissions cutoffs. For example, \(r_{\lambda,t+1}\), \(K_{t+1}\), and \(S_{t+1}^*\) could depend on placements, research output, and aggregate policies. The baseline does not impose these extra laws of motion because they are not needed for the production--evaluation mechanism studied in the main text.

\section{Reduced-Form Welfare Accounting}
\label{app:welfare}

The baseline model characterizes private incentives. This section gives a reduced-form welfare accounting to separate private ranking gains from broader social value. The object below is not a welfare theorem; it is a compact way to organize the main forces in the model.

Let \(q\) denote accumulated research capability, distinct from the fixed ability type \(\theta\). For an RA in a type-\(\lambda\) laboratory, define expected capability growth as
\[
\Delta q_\lambda(\theta)
\equiv
\mathbb{E}[q'-q\mid \theta,\lambda].
\]
In the learning extension, when the learning shock has mean zero,
\[
\Delta q_\lambda(\theta)
=
\ell\!\left(e_\lambda^*(\theta),K_{N,\lambda}^T\right)
+
\psi_L\log(1+g_\lambda).
\]

A representative type-\(\lambda\) reduced-form welfare object, expressed per filled position with segment-scale costs allocated to the representative position, is
\begin{equation}
W_\lambda^S
=
\mathbb{E}_\lambda
\left[
\omega_Q Y_\lambda(\theta)
+
\omega_H \Delta q_\lambda(\theta)
-
\frac{(e_\lambda^*(\theta))^2}
{2\chi(\theta,\alpha_\lambda K)}
\right]
-
\frac{c_g}{2}g_\lambda^2
-
\frac{c_n}{2}n_\lambda,
\label{eq:welfare_appendix}
\end{equation}
where \(\omega_Q>0\) is the social value of knowledge output and \(\omega_H>0\) is the social value of human-capital accumulation. The expectation is over the RA types filling positions in segment \(\lambda\). The last term is the per-position allocation of the segment-level team-intensity cost \((c_n/2)n_\lambda^2\). A full-market welfare measure would aggregate these segment objects as \(\sum_\lambda \mu_\lambda n_\lambda W_\lambda^S\). Because \(K\) is treated as an external AI frontier in the baseline, this accounting does not include a separate AI-adoption cost.

This expression shows why the welfare effect of AI is not signed by the baseline mechanism alone. AI can raise social value by increasing research output and future capability \citep{becker1962investment,manso2011incentives}. But in a fixed-capacity tournament, some private gains from stronger visible records may be converted into higher cutoffs, extra effort, and signal escalation rather than proportional social value \citep{lazear1981rank,hopkins2023is}. The welfare effect is positive when output and learning gains dominate, and weaker when congestion and induced resource costs dominate.

\section{Project-Level Research Uncertainty and Breakthrough Payoff}
\label{app:uncertainty}

This section supports the breakthrough component in the quality-oriented PI payoff. The baseline separates two kinds of uncertainty. Signal noise, \((\varepsilon_R,\varepsilon_N)\), affects what the admissions market observes. Project-level uncertainty, \(\zeta\), affects whether a novel research project ultimately crosses a high-impact threshold. This lets the model treat exploratory research as risky without turning the baseline into a full stochastic-discovery model \citep{just1978stochastic, aghion2008academic, manso2011incentives}.

Let realized novel-project merit be
\begin{equation}
q_N
=
y_N+\zeta,
\qquad
\zeta\sim N(0,\sigma_\zeta^2),
\label{eq:qN_project_uncertainty}
\end{equation}
where \(\zeta\) is independent of \((\theta,\varepsilon_R,\varepsilon_N)\). A breakthrough occurs when
\begin{equation}
q_N>\bar q.
\label{eq:breakthrough_project_uncertainty}
\end{equation}
Since
\[
y_N
=
\kappa_\lambda(K_N^T)(\theta+\eta_N e+\psi g),
\qquad
K_N^T=(1-\alpha)K,
\]
the conditional breakthrough probability is
\begin{equation}
p_{B,\lambda}^{\zeta}(\theta,e,g,\alpha,K)
=
1-
\Phi\!\left(
\frac{
\bar q-
\kappa_\lambda(K_N^T)(\theta+\eta_N e+\psi g)
}{
\sigma_\zeta
}
\right).
\label{eq:pB_project_uncertainty}
\end{equation}
The segment-level breakthrough probability used in the PI payoff is
\begin{equation}
\bar p_{B,\lambda}
=
\int_{\Theta_\lambda}
p_{B,\lambda}^{\zeta}
\bigl(\theta,e_\lambda^*(\theta),g_\lambda,\alpha_\lambda,K\bigr)
\,dF_\lambda(\theta),
\label{eq:avg_breakthrough_appendix}
\end{equation}
where \(F_\lambda\) is the conditional distribution of RAs filling positions in segment \(\lambda\).

To see the relevant comparative statics, define the deterministic novel-task component
\[
z_N
=
\kappa_\lambda(K_N^T)(\theta+\eta_N e+\psi g).
\]
Because
\[
\frac{\partial}{\partial z_N}
\left[
1-\Phi\!\left(\frac{\bar q-z_N}{\sigma_\zeta}\right)
\right]
=
\frac{1}{\sigma_\zeta}
\varphi\!\left(\frac{\bar q-z_N}{\sigma_\zeta}\right)
>0,
\]
the breakthrough probability increases with the deterministic novel-task component. Hence, holding the other arguments fixed,
\[
\frac{\partial p_{B,\lambda}^{\zeta}}{\partial K_N^T}
=
\frac{1}{\sigma_\zeta}
\varphi\!\left(\frac{\bar q-z_N}{\sigma_\zeta}\right)
\kappa_\lambda'(K_N^T)(\theta+\eta_N e+\psi g)
>0,
\]
and
\[
\frac{\partial p_{B,\lambda}^{\zeta}}{\partial g}
=
\frac{1}{\sigma_\zeta}
\varphi\!\left(\frac{\bar q-z_N}{\sigma_\zeta}\right)
\kappa_\lambda(K_N^T)\psi
>0.
\]
Since \(K_N^T=(1-\alpha)K\), an increase in the routine-task AI share \(\alpha\) reduces breakthrough probability through the novel-task AI channel:
\[
\frac{\partial p_{B,\lambda}^{\zeta}}{\partial \alpha}
=
-K
\frac{\partial p_{B,\lambda}^{\zeta}}{\partial K_N^T}
<0
\qquad
(K>0).
\]

Thus, mentoring and novel-task AI raise the probability of upper-tail research success, while shifting AI capacity toward routine tasks lowers it. This is the project-risk channel behind the quality-oriented PI's incentive to preserve mentoring and novel-task augmentation. Because \(\zeta\) is distinct from \((\varepsilon_R,\varepsilon_N)\), this risk is discovery uncertainty, not admissions measurement noise.

\section{Simulation Parameters and Supplementary Figures}
\label{app:simulation_parameters}

The simulation in Section~5 is illustrative rather than structurally estimated. Table~\ref{tab:simulation_parameters_appendix} reports the parameter values used in the enhanced run. The purpose is reproducibility and scale transparency: the exercise asks whether the paper's mechanisms can coexist in a dense agent-level environment with distributed talent, continuous PI orientation, project luck, reception noise, and fixed capacity.

\begin{table}[h!]
\centering
\caption{Enhanced simulation parameters}
\label{tab:simulation_parameters_appendix}
\begin{adjustbox}{width=\linewidth}
\begin{tabular}{lll}
\toprule
Block & Object & Value\\
\midrule
Population & RAs per replication & \(I=4{,}000\)\\
Population & PIs per replication & \(J=200\)\\
Population & Monte Carlo replications & \(80\) per main scenario--\(K\) point\\
Population & AI frontier grid & \(K=0.00,0.10,\ldots,2.00\)\\
Population & Capacity ratio & \(Q/M=0.170\)\\
Population & Continuous PI grid & \(31\) \(\lambda\)-grid points for policy precomputation\\
\midrule
RA talent & Mean routine talent & \(\mu_{\theta^R}=1.00\)\\
RA talent & Mean novel talent & \(\mu_{\theta^N}=1.00\)\\
RA talent & Talent standard deviations & \(\sigma_{\theta^R}=0.22,\;\sigma_{\theta^N}=0.25\)\\
RA talent & Talent correlation & \(\rho_{\theta}=0.55\)\\
RA talent & Truncation bounds & \([\theta_L,\theta_H]=[0.20,1.90]\)\\
\midrule
Effort and input & Effort capacity & \(\chi(\theta^L,x)=0.32+0.22\theta^L+0.16x\)\\
Effort and input & Effort return scale & \(\beta_{RA}=0.95\)\\
Effort and input & Effort loadings & \(\eta_R=0.35,\;\eta_N=0.45\)\\
Effort and input & Mentoring loading & \(\psi=0.30\)\\
\midrule
Signals and uncertainty & Routine signal noise & \(\sigma_R=0.20\)\\
Signals and uncertainty & Novel signal noise & \(\sigma_N=0.28\)\\
Signals and uncertainty & Project-luck schedule & \(\sigma_\zeta(\lambda)=0.22(1+0.85\lambda)\)\\
Signals and uncertainty & Reception-noise schedule & \(\sigma_\xi(\lambda)=0.10(1+0.65\lambda)\)\\
Signals and uncertainty & Diagnostic compression & \(\rho_R(x)=\exp(-\delta_\rho x),\;\delta_\rho=1.10\)\\
\midrule
Production & Routine automation level & \(a_R(x)=0.10+1.05x\)\\
Production & Routine human multiplier & \(m_R(x)=1.00+0.12x\)\\
Production & CES routine share & \(\tau=0.68\)\\
Production & Breakthrough threshold & \(\bar q=2.45\)\\
\midrule
PI endpoints & \(\phi_\lambda\) & quantity \(0.16\), quality \(0.72\)\\
PI endpoints & \(\varsigma_\lambda\) & quantity \(1.90\), quality \(0.72\)\\
PI endpoints & \(r_\lambda\) & quantity \(0.98\), quality \(1.15\)\\
PI endpoints & \(\Omega_\lambda\) & quantity \(0.00\), quality \(3.00\)\\
PI endpoints & \(\gamma_\lambda\) & quantity \(1.12\), quality \(0.20\)\\
\midrule
Cost and policy grid & Wage and cost terms & \(\bar w=0.58,\;c_g=0.55,\;c_n=0.45\)\\
Cost and policy grid & Routine AI share grid & \(\alpha\in[0.05,0.95]\), 19 points\\
Cost and policy grid & Mentoring grid & \(g\in[0.00,2.20]\), 12 points\\
Cost and policy grid & Filled team-intensity grid & \(n\in[0.40,3.20]\), 12 points\\
\bottomrule
\end{tabular}
\end{adjustbox}
\end{table}

Table~\ref{tab:simulation_sensitivity_appendix} summarizes the main sensitivity checks. These checks are not alternative calibrations. They show that the qualitative patterns emphasized in Section~5 do not rely on a single capacity ratio, noise scale, diagnostic-compression strength, or coarse \(\lambda\)-grid. Table~\ref{tab:simulation_effort_assignment_appendix} then addresses two implementation concerns directly: the baseline effort-loading approximation and the filled-random RA assignment rule.

\begin{table}[h!]
\centering
\caption{Simulation sensitivity checks}
\label{tab:simulation_sensitivity_appendix}
\begin{adjustbox}{width=\linewidth}
\begin{tabular}{lll}
\toprule
Check & Statistic & Finding\\
\midrule
Capacity ratio, \(Q/M\in\{0.10,0.17,0.25,0.35\}\) at \(K=2\) &
\(FN^{ability}\), cutoff &
\makecell[l]{Capacity expansion lowers \(FN^{ability}\) from \(0.741\) to \(0.375\);\\
the cutoff falls from \(1.491\) to \(1.242\).}\\
\midrule
Assessment-noise scale, \(0\) to \(4\), holding \(K=0\) &
\(FN^{ability}\), \(\mathrm{corr}(S^{obs},M^{true})\) &
\makecell[l]{Higher reception noise raises \(FN^{ability}\) from \(0.467\) to \(0.725\);\\
score-merit alignment falls from \(0.445\) to \(0.155\).}\\
\midrule
Project-luck scale, \(0.25\) to \(3\), holding \(K=0\) &
\(FN^{ability}\), \(\mathrm{corr}(S^{obs},M^{true})\) &
\makecell[l]{Ability false negatives remain around \(0.54\)--\(0.55\);\\
score-merit alignment falls from \(0.603\) to \(0.149\).}\\
\midrule
Diagnostic-compression strength, \(\delta_\rho\in\{0.70,1.10,1.50\}\) at \(K=2\) &
Precision, \(FN^{ability}\) &
\makecell[l]{Routine diagnostic precision falls from \(5.657\) to \(2.766\);\\
\(FN^{ability}\) ranges from \(0.611\) to \(0.623\).}\\
\midrule
Finer continuous-orientation grid, \(61\) \(\lambda\)-points at \(K=2\) &
\(\mathrm{corr}(\lambda,\alpha)\), \(\mathrm{corr}(\lambda,g)\), \(\mathrm{corr}(\lambda,n)\) &
\makecell[l]{The segmentation pattern is unchanged:\\
\(-0.939\), \(0.846\), and \(-0.960\), respectively.}\\
\bottomrule
\end{tabular}
\end{adjustbox}
\end{table}

Table~\ref{tab:simulation_effort_assignment_appendix} reports a stronger robustness exercise for the two simulation choices that are furthest from the analytical equilibrium. First, the cutoff-sensitive effort variant replaces the baseline loading rule with a one-dimensional best response: given a simulated cutoff, each RA chooses effort to maximize expected admission value net of quadratic effort cost, and the cutoff is updated from the induced score distribution. Second, the rank-sorted assignment variant preserves the same filled team-intensity masses but assigns RAs by comparative novel-versus-routine talent, \(\theta_i^N-\theta_i^R\), across the PI-orientation ranking. These are deliberately conservative checks. The first moves effort closer to the cutoff-dependent RA problem in the analytical model; the second gives the simulation a strong sorting structure rather than random assignment across filled positions.

\begin{table}[h!]
\centering
\caption{Effort-rule and assignment-rule robustness at \(K=2\)}
\label{tab:simulation_effort_assignment_appendix}
\begin{adjustbox}{width=\linewidth}
\begin{tabular}{lccccc}
\toprule
Variant & Mean effort & \(\mathrm{corr}(S^{obs},\theta^L)\) & \(FN^{ability}\) & \(FN^{merit}\) & Interpretation\\
\midrule
\makecell[l]{Baseline loading effort;\\filled-random assignment} &
0.175 & 0.510 & 0.617 & 0.838 &
\makecell[l]{Main simulation\\specification}\\
\midrule
\makecell[l]{Cutoff-sensitive effort;\\filled-random assignment} &
0.151 & 0.534 & 0.620 & 0.821 &
\makecell[l]{Cutoff feedback leaves\\ability false negatives similar}\\
\midrule
\makecell[l]{Loading effort;\\rank-sorted assignment} &
0.176 & 0.513 & 0.608 & 0.887 &
\makecell[l]{Ability sorting does not\\remove realized-merit errors}\\
\midrule
\makecell[l]{Cutoff-sensitive effort;\\rank-sorted assignment} &
0.130 & 0.518 & 0.618 & 0.871 &
\makecell[l]{Both stronger checks\\preserve the main pattern}\\
\bottomrule
\end{tabular}
\end{adjustbox}
\begin{minipage}{0.96\linewidth}
\footnotesize
Notes: All rows use the diagnostic-compression environment at \(K=2\), \(Q/M=0.17\), and the same PI policy precomputation. The cutoff-sensitive effort rule is a robustness approximation, not a replacement for the analytical effort problem: it solves a scalar effort best response against the simulated cutoff and iterates the cutoff a small number of times. The rank-sorted assignment rule preserves the filled team-intensity masses \(n_\lambda\) but assigns higher comparative novel talent to higher-\(\lambda\) PI environments. The table shows that the main selection-error result is not driven by the baseline effort-loading approximation or by filled-random assignment.
\end{minipage}
\end{table}

\begin{figure}[h!]
\centering
\includegraphics[width=\linewidth]{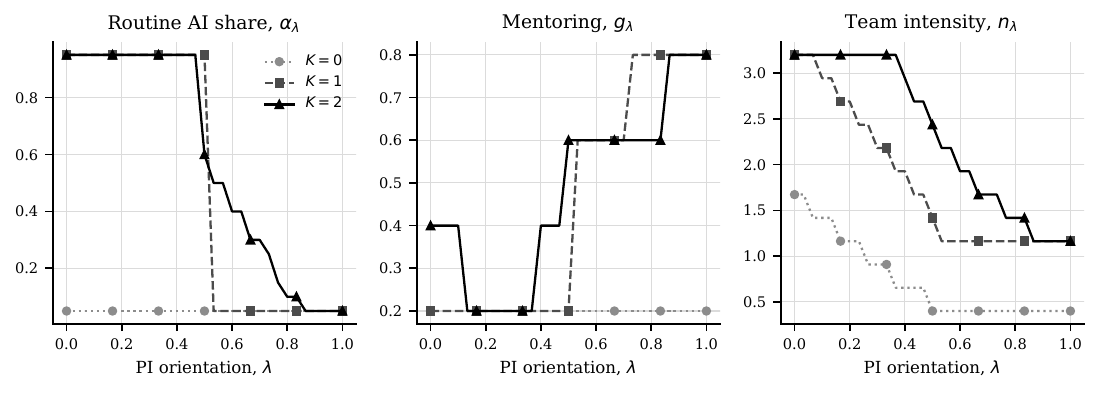}
\caption{Continuous PI orientation and organizational choices. The horizontal axis is the continuous PI-orientation index \(\lambda\), with lower values corresponding to the quantity-oriented endpoint and higher values corresponding to the quality-oriented endpoint. Curves report the diagnostic-compression environment at \(K=0\), \(K=1\), and \(K=2\). Panel A plots the routine-task AI share \(\alpha_\lambda\); higher values mean that a larger share of AI capacity is allocated to routine tasks rather than novel tasks. The \(K=0\) line in Panel A is included only as the grid baseline, because there is no positive AI capacity to allocate when \(K=0\); substantive interpretation of AI allocation focuses on \(K>0\). Panel B plots mentoring intensity \(g_\lambda\), the PI investment that raises the novel-task human input channel. Panel C plots filled team intensity \(n_\lambda\), the mass of RA positions operated by a PI after participation and sorting. The figure shows that the two-type analytical segmentation is an endpoint representation of a continuum: lower-\(\lambda\) PIs rely more on routine-task automation and larger teams, while higher-\(\lambda\) PIs preserve more mentoring and less routine-task automation. The stepwise shape reflects grid-based policy optimization and threshold-like organizational regimes; Table~\ref{tab:simulation_sensitivity_appendix} reports a finer-grid check. The figure is illustrative and is not a structural estimate.}
\label{fig:abm_orientation_appendix}
\end{figure}

Fig.~\ref{fig:abm_orientation_appendix} reports the continuous-orientation organizational policies that are summarized in the main text. The figure is placed here because it supports the robustness of the segmentation mechanism rather than the main simulation's ability--luck selection result. The stepwise shape reflects grid-based policy optimization and threshold-like organizational regimes; the 61-point grid check in Table~\ref{tab:simulation_sensitivity_appendix} verifies that the monotone segmentation pattern is not an artifact of the baseline grid.

\begin{figure}[h!]
\centering
\includegraphics[width=\linewidth]{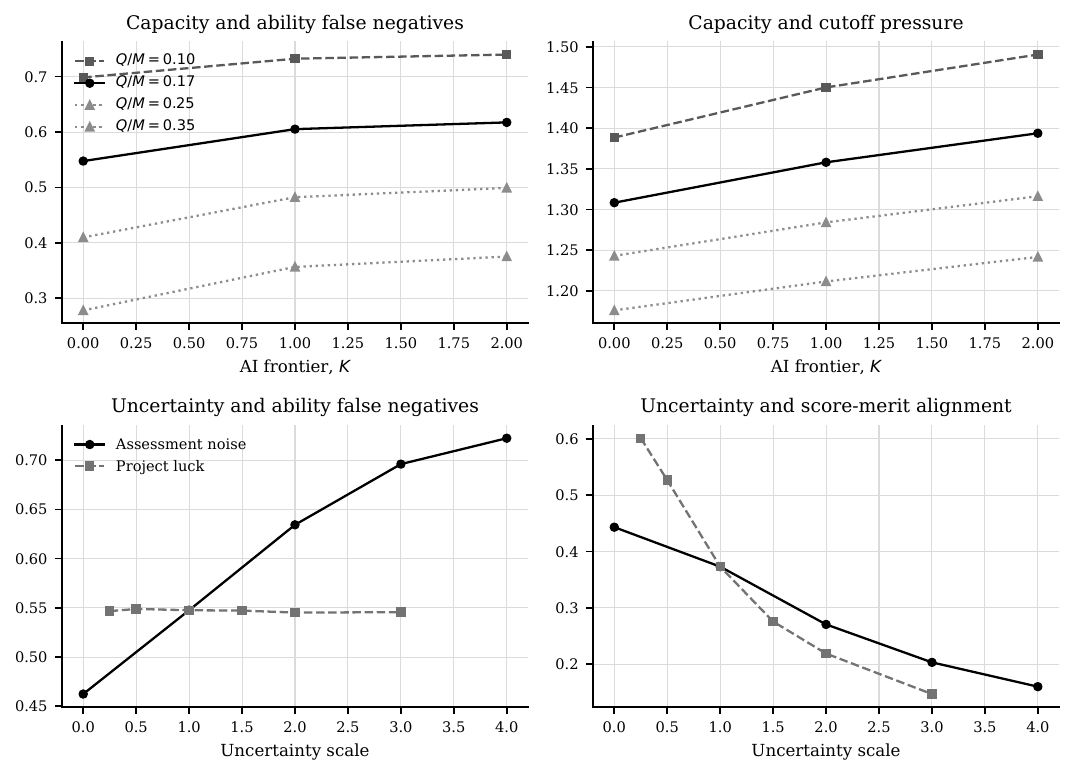}
\caption{Bottlenecks, luck, and assessment uncertainty. Panels A and B vary the admissions capacity ratio \(Q/M\) in the diagnostic-compression environment. A lower \(Q/M\) means a tighter bottleneck: fewer candidates can be admitted for a given mass of participating RAs. Panel A plots ability-based false negatives, \(FN^{ability}\), defined as the probability that a candidate is rejected conditional on latent ability \(\theta_i^L\) being in the top quintile. Panel B plots the admissions cutoff \(S^*\) over the same \(K\) grid, showing how capacity changes both rejection risk and cutoff pressure. Panels C and D hold \(K=0\) so that uncertainty is isolated before AI diagnostic compression is introduced. They vary a multiplicative uncertainty scale applied either to reception noise \(\xi\), which enters the perceived score \(S^{\mathrm{obs}}\), or to project luck \(\zeta\), which enters realized merit \(M^{\mathrm{true}}\). Panel C shows that reception noise directly raises ability-based false negatives, while project luck has weaker direct effects because admissions do not observe \(\zeta\). Panel D shows that both kinds of uncertainty weaken score-merit alignment, especially project luck because it drives realized merit away from latent ability and evaluated evidence. The figure is illustrative and is not a structural estimate.}
\label{fig:abm_bottleneck_uncertainty_appendix}
\end{figure}

Fig.~\ref{fig:abm_bottleneck_uncertainty_appendix} reports the bottleneck and uncertainty sensitivity checks summarized in the main text. These panels complement Table~\ref{tab:simulation_sensitivity_appendix}: capacity expansion mainly lowers ability-based false negatives, reception noise worsens the mapping from score to latent ability, and project luck widens the gap between latent ability and realized project merit.

\clearpage
\processdelayedfloats
\clearpage

\bibliography{references}